\pdfoutput=1
\documentclass[aps,preprint,amsmath,amsfonts,amssymb,eqsecnum,nofootinbib,
  floatfix,secnumarabic,preprintnumbers,superscriptaddress,nobalancelastpage,
  onecolumn,titlepage]{revtex4}
\usepackage{color,graphicx,slashed,hyperref}
\usepackage[caption=false]{subfig}
\hypersetup{
   unicode=true,          % non-Latin characters in Acrobat�s bookmarks
   pdftoolbar=true,        % show Acrobat�s toolbar?
   pdfmenubar=true,        % show Acrobat�s menu?
   pdffitwindow=false,     % window fit to page when opened
   pdfstartview={FitH},    % fits the width of the page to the window
   pdftitle={Light fermion DM},    % title
   pdfauthor={Author},     % author
   pdfsubject={Subject},   % subject of the document
   pdfcreator={Creator},   % creator of the document
   pdfproducer={Producer}, % producer of the document
   pdfkeywords={keyword1} {key2} {key3}, % list of keywords
   pdfnewwindow=true,      % links in new PDF window
   colorlinks=true,       % false: boxed links; true: colored links
   linkcolor=black,          % color of internal links (change box color with linkbordercolor)
   citecolor=magenta,        % color of links to bibliography
   filecolor=magenta,      % color of file links
   urlcolor=black           % color of external links
}

\definecolor{red}{rgb}{1.0, 0, 0}
\definecolor{orange}{rgb}{1,0.,1}
\newcommand{\be}{\begin{equation}}
\newcommand{\ee}{\end{equation}}
\def\bsp#1\esp{\begin{split}#1\end{split}}

\def\nn{\nonumber}
\def\piz{\ensuremath{{\pi^0}}}

\def\udark{U(1)_D}
\def\gzp{g_{V}}
\def\aem{\alpha_{\rm em}}
\def\azp{\alpha_{D}}
\def\ydm{y_{\textrm{DM}}}

\def\zp{V{}}
\def\mzp{M_{V}}
\def\ms{M_{S}}
\def\mdm{M_{\chi}}
\def\mdone{M_{\chi_1}}
\def\mhds{M_{\chi_2}}
\def\dchi{\Delta_{\chi}}
\def\eps{\varepsilon}
\def\ysr{y_{R}}
\def\ysl{y_{L}}

\begin{document}
%\date{\today}

\title{\vspace{1cm} Signatures of dark Higgs boson \\ in light fermionic dark matter scenarios\vspace{1cm}  }

\author{Luc Darm\'e}
\email{luc.darme@ncbj.gov.pl}
\affiliation{National Centre for Nuclear Research, ul. Ho\.za 69, 00-681 Warsaw, Poland}

\author{Soumya Rao}
\email{soumya.rao@ncbj.gov.pl}
\affiliation{National Centre for Nuclear Research, ul. Ho\.za 69, 00-681 Warsaw, Poland}

\author{Leszek Roszkowski}
\email{leszek.roszkowski@ncbj.gov.pl}
\affiliation{Astrocent, Nicolaus Copernicus Astronomical Center Polish Academy of
	Sciences, ul. Bartycka 18, 00-716 Warsaw, Poland\vspace{1cm} } 
\affiliation{National Centre for Nuclear Research, ul. Ho\.za 69, 00-681 Warsaw, Poland}

% \pacs{}
%\preprint{}

\begin{abstract}
Thermal dark matter scenarios based on light (sub-GeV) fermions typically require the
presence of an extra dark sector containing both a massive dark photon along with a dark
Higgs boson. The latter typically generates both the dark photon mass and an additional mass term for the dark
sector fermions. This simple setup has both rich phenomenology and
bright detection prospects at high-intensity accelerator experiments. We point out that in
addition to the well studied pseudo-Dirac regime, this model can achieve the correct relic
density in three different scenarios, and examine in details their properties and experimental prospects. We emphasize in
particular the effect of the dark Higgs boson on both detection prospects and cosmological
bounds. 
%In a large part of the parameter space, the dark Higgs boson appears to be lightest degrees of freedom of the theory. This setup will present both a strong signature in beam dump experiments along with various Higgs Sommerfeld enhancements.
\end{abstract}

\maketitle

\section{Introduction}\label{intro}

Weakly interacting massive particle (WIMP) as a dark matter (DM) candidate with a sub-GeV
mass scale has attracted a great deal of attention (for recent reviews see
e.g.\cite{Roszkowski:2017nbc, Plehn:2017fdg, Arcadi:2017kky}).  Much recent effort has
gone to explore the nature and properties of DM of sub-GeV mass interacting through a
light mediator of the same scale, hence reproducing WIMP scenarios at the MeV scale.  Such
scenarios involve the presence of a MeV scale vector boson charged under an extra
$\udark$, usually referred to as dark photon, that acts as the light mediator for dark
matter interactions.  The kinetic mixing between the dark photon and the Standard Model
(SM) photon is then invoked to introduce interactions between the dark sector and SM
particles (see \cite{Alexander:2016aln, Battaglieri:2017aum} for reviews). We focus on a
model containing a complete, self consistent ``dark sector'' built from a Dirac fermion
DM candidate, a dark photon and a dark Higgs boson. Interestingly, bounds from
cosmic microwave background (CMB) observations can be avoided by introducing Yukawa
couplings between the dark Higgs boson and the DM field which lead to an
additional Majorana mass term once the dark Higgs boson acquires its vacuum expectation
value. 

In this article, we build on our previous work~\cite{Darme:2017glc} and show that this
model exhibits a very rich DM phenomenology. Indeed, we have identified four
distinct scenarios which achieve the correct relic density while avoiding the bounds from
CMB: (i) an inelastic DM ($iDM$) regime, (ii) a Majorana DM ($mDM$) regime, (iii) a
secluded regime and (iv) a forbidden regime. The $iDM$ regime is defined as the region of
the parameter space with pseudo-Dirac DM composed of two Majorana states with a
small mass splitting.  This scenario has been well studied in the past, and the dark Higgs
boson does not play any significant role there.  The $mDM$ regime has larger mass splitting
between the two Majorana mass eigenstates, of the order of the DM mass, and markedly
different detection prospects in accelerator-based experiments.  The secluded regime
refers to the region where the DM mass is close to the dark Higgs boson mass such
that the annihilation channel into dark Higgs bosons becomes critical in reaching the
thermal target.  Note that this definition of ``secluded'' region is different from the
one used in previous literature \cite{Batell:2009yf}.  And finally, the forbidden regime is the region of
parameter space where the DM mass is close to that of the dark photon but less than that
of the dark Higgs boson.  Thus, the annihilation channel of DM into dark photon pair or a
dark Higgs boson and a dark photon is forbidden kinematically, but nonetheless contribute
to relic density due to thermal effects.

While direct detection experiments are not typically sensitive to our model due to various
suppression mechanisms~\cite{Battaglieri:2017aum} (since scattering processes are either
inelastic or velocity-suppressed), detection prospects in accelerator-based experiments
are bright and have been enthusiastically studied in recent years for a variety of
contexts and experimental projects (some very recent examples are for
instance~\cite{Correia:2016xcs,Knapen:2017xzo,Feng:2017drg,Feng:2017vli,Wojtsekhowski:2017ijn,Feng:2017uoz,Berlin:2018pwi,Berlin:2018bsc,Jordan:2018gcd}).
However, bounds for sub-GeV fermionic DM were mainly obtained in the pseudo-Dirac regime,
either by searching for the decay of the long-lived heavy component of the pseudo-Dirac
DM, or by focusing instead on the signatures from DM scattering, as advertised in,
e.g.~\cite{Batell:2009di,Batell:2014mga,Izaguirre:2015yja,deNiverville:2016rqh,Battaglieri:2016ggd}
(a important exception is of course search focusing on the dark photon itself, which
remain relatively agnostic about the details of the dark sector, see
\cite{Alexander:2016aln,Battaglieri:2017aum} for a review). 

Our main point is that the accelerator-based bounds differ significantly in all of the
four regimes identified above, especially when the parameter space is restricted to the
parameter space reaching the thermal target. In order to illustrate this fact, we focus on
two old experiments: LSND \cite{Athanassopoulos:1996ds} that uses a proton beam and a
relatively short beamline and the electron beam dump experiment E137
\cite{Riordan:1987aw},  and recast their null results into constraints for each of the four
regimes. On top of providing the current experimental status of each regime, this shows
that the pseudo-Dirac regime is not the only region of the parameter space which can be
significantly constrained by accelerator experiments. Furthermore, we show that when
long-lived, the dark Higgs boson can significantly alter the accelerator phenomenology of
our model, by either providing sizable new constraints (mainly in the forbidden regime),
or on the contrary by reducing dramatically their reach. In particular, while searches for
dark Higgs boson had already been advertised long ago in, e.g.
\cite{Batell:2009di,Bjorken:2009mm,Schuster:2009au,Essig:2009nc} and considered in our
previous work~\cite{Darme:2017glc}, we complement these analyses by including the dark
Higgs boson decay into a dark photon and an $e^+e^-$ pair. 

%Finally, we further expand~\cite{Darme:2017glc} by considering constraints on light dark sector particles in the MeV range from
%supernova 1987A (SN1987A) \cite{Burrows:1986me, Burrows:1987zz, Raffelt:1987yt,
%Raffelt:1996wa} have been studied in several articles \cite{Bjorken:2009mm, Dent:2012mx,
%An:2013yfc, Redondo:2013lna, Kazanas:2014mca, Rrapaj:2015wgs, Chang:2016ntp,
%Hardy:2016kme, Mahoney:2017jqk, Chang:2018rso}.  We briefly comment about the relevance of
%the constraints obtained in \cite{Chang:2018rso} for an inelastic DM model that
%is similar to the model studied here.

The paper is organized as follows. We begin by describing the light fermionic DM model
studied here in Sec.~\ref{sec:model}.  We also describe the different regions of parameter
space mentioned above in more detail, and briefly review the cosmological constraints on
the dark Higgs boson.  In Sec.~\ref{sec:bdump}, we present the constraints from beam dump
experiments like LSND and E137 for each of the scenarios mentioned above.  In
Sec.~\ref{sec:res} we summarize our results and conclude.

\section{Building a light fermionic DM model}\label{sec:model}

\subsection{A self-consistent minimal framework}
\label{sec:modbuilding}

Our goal is to build a model of thermal fermionic DM with mass in the sub-GeV range while
relying essentially on simple SM-like building blocks. Arguably the simplest and least
constrained way of doing this is to rely on a vector mediator mechanism, where the portal
between the dark sector and the SM is provided by the kinetic mixing between a new,
spontaneously-broken, abelian gauge group $\udark$ (under which the SM is
neutral)\footnote{Note that building models where part of the SM fields are charged under
the new dark gauge group is possible but non-trivial, see
e.g.~\cite{Ellis:2017tkh}.} and the SM $U(1)_Y$. In this approach,
two main constraints need to be factored in: gauge anomaly cancellation
for the dark gauge group, and indirect detection bounds. The former implies that
at least two fermionic fields with opposite $\udark$ charge must be added to the
theory, and the latter that their mass terms must be at least partially of
Majorana type to avoid an $s$-wave annihilation through an off-shell dark photon.
Interestingly, both constraints can be straightforwardly satisfied by including a
Yukawa coupling between the dark Higgs boson and these new fermionic states.

More precisely, we define  a Dirac fermion $\chi = (\chi_L,\bar{\chi}_R)$ DM with charge
$1$ under the dark $U(1)$, which will acquire additional Majorana masses from its Yukawa
interactions with the dark Higgs boson of $\udark$ charge $q_S = +2$. The effective
Lagrangian for the dark photon vector ($V$) and the dark Higgs boson ($S$) fields in this
minimal dark sector model is then given by
\begin{align}
%\mathcal{L_{eff}}=\mathcal{L}_{int}+\mathcal{L}_{DM}
\mathcal{L}_{V} &= -\frac{1}{4}F^{\prime\mu\nu}F^{\prime}_{\mu\nu}
-\frac{1}{2}\frac{\varepsilon}{\cos\theta_w}B_{\mu\nu}F^{\prime\mu\nu} \ , \\ 
\mathcal{L}_S&=(D^\mu S )^*(D_\mu S)+ \mu_S^2 |S|^2 - \frac{\lambda_S}{2} |S|^4 - \frac{\lambda_{SH}}{2} |S|^2  |H|^2 \ , 
\label{lma}
\end{align}
where $F^{\prime}\mu\nu$ is the corresponding field tensor for $\udark$, and we have
introduced a kinetic mixing term parametrized by $\eps$. This term is an
invariant of both gauge groups which, even if not present at tree-level in the theory, can
be generated by loop corrections of heavy vector-like fermion charged under both
$U(1)$s~\cite{Holdom:1985ag}. Furthermore, we will fix the value of the gauge-preserving quartic
coupling mixing the dark Higgs boson with the SM Higgs to zero, since, as was shown
in~\cite{Darme:2017glc}, its effects are negligible for values compatible with a natural
splitting between dark sector and the SM (see also~\cite{morrissey} where this coupling is
generated directly by supersymmetry, as is shown to have small values).
The DM field is then introduced through the Lagrangian
\begin{align}
\mathcal{L}^{\rm DM}&=\bar{\chi}\left( i\slashed{D}-m_{\chi}\right)\chi + \left( \ysl S \bar{\chi}^c P_L \chi + \ysr S \bar{\chi}^c P_R  \chi_R + \textrm{ h.c.} \right) \,  \ .
\end{align}
After the  dark Higgs boson acquires a Vacuum Expectation Value (VEV), the mass of the
dark Higgs boson and of the dark photon become correlated and are given by
\begin{align}
\label{eq:mass}
M_S &= \sqrt{2 \lambda_S} v_S \ , \\
M_{\zp} &= \gzp q_S v_S = \left( \frac{q_S \gzp }{\sqrt{2 \lambda_S}} \right) M_S\ .
\end{align}
In term of Weyl fermions, the interaction and mass terms for the DM fields are
\begin{align}
\mathcal{L}^{\rm DM} \supset~ & \left(\frac{1}{2} (\chi_L,\chi_R)  ~M_\chi \begin{pmatrix}
\chi_L\\
\chi_R 
\end{pmatrix} + \frac{\ysl}{\sqrt{2}} S  \chi_L \chi_L + \frac{\ysr}{\sqrt{2}} S  \bar{\chi}_R \bar{\chi}_R +h.c\right) \nonumber \\
 &+ \bar{\chi}_R \bar{\sigma}^\mu \zp_{\mu} \chi_R - \bar{\chi}_L \bar{\sigma}^\mu \zp_{\mu} \chi_L \ ,
\end{align}
where the DM mass matrix is 
\begin{align}
 M_\chi = \begin{pmatrix}
           \sqrt{2} v_S \ysl & m_\chi \\
           m_\chi & \sqrt{2} v_S \ysr 
          \end{pmatrix} \ .
\end{align}
The lightest eigenstate has typically a negative mass and the splitting between the two mass eigenstates $\chi_1$ and $\chi_2$ is
\begin{align}
 \dchi \equiv M_{\chi_1} + M_{\chi_2} = \sqrt{2 } \frac{\mzp}{\gzp} (\ysr+\ysl ) \ .
\end{align}
As usual in this case, there are typically two possible conventions, one can either keep
the negative mass but ensure that the rotation matrix in the DM sector $Z^X$ is real, or
else ensure that both eigenstates have positive masses, at the expense of introducing an
imaginary rotation matrix. Following~\cite{TuckerSmith:2001hy}, most authors considering
the pseudo-Dirac limit have used the latter choice as this leads to a very simple form for
the gauge interaction between mass eigenstates $\chi_1$ and $\chi_2$. We will make
throughout this paper the opposite choice to have potentially a negative mass for $\chi_1$
but real rotation matrices as this is easier to implement numerically while not in the
pseudo-Dirac limit.
In the following, we will also denote by $\chi_1$ and $\chi_2$ the Majorana fermions
corresponding to the previously defined Weyl fermion states for notational simplicity. We also use the notation $\mdm$ (e.g in plots) to refer to the absolute value of the DM mass $\mdm = | \mdone|$ and when $\mdone > 0$, we define $\dchi \equiv \mhds - \mdone$.

Defining the effective gauge interaction between $\chi_1$ and $\chi_2$ by $g_{D, 12}$,
we have
\begin{align}
g_{D, 12} = \gzp ~ \frac{1}{\sqrt{1 + \frac{\mzp^2 (\ysl-\ysr)^2}{2 \gzp^2 m_\chi^2}}}  \ ,
 \label{eq:iDMcoupling}
\end{align}
which reduces to the dark gauge coupling $\gzp$ when the Dirac mass dominates, as expected. On the other hand, the DM candidate $\chi_1$ has an effective gauge coupling given by
\begin{align}
 g_{D, 11} = \gzp ~ \frac{\mzp (\ysl-\ysr)}{\sqrt{2 \gzp^2 m_\chi^2 + \mzp^2
 (\ysl-\ysr)^2}},  
 \label{eq:mDMcoupling}
\end{align}
which vanishes in the limit where % should this not be just ysl=ysr? and then say that it approximately becomes zero(or can be neglected) for ysl~ysr and ysl,ysr<<1?
$\ysl = \ysr$ and is very small for $\ysl,\ysr \ll 1$ or $\ysl \sim \ysr$.
%  \qquad \qquad y_{DM} = (\ysl+\ysr) -(\ysl-\ysr) \frac{g_{DM}}{\gzp}
%\LD{Check these formula}

A key element of this model is the fact that there are only two independent mass scales:
the fermion Dirac mass and the dark Higgs boson VEV whose interplay will determine most of
the phenomenology of the model. As we will see later, the DM mass is typically restricted
by CMB bounds and relic density evaluation to be smaller than the dark photon one: $\mdm <
\mzp$. Freedom in the spectrum of the model is therefore mainly determined by the position of
the dark Higgs boson mass and the four regimes we will identify below largely depends on it. 
%The main ordering freedom in the spectrum of the model is therefore the position of the dark Higgs boson mass, and the four regimes you will consider below largely depends on it.

Analytically, we find that the dark Higgs boson is lighter than the dark photon when
\begin{align*}
\sqrt{2 \lambda_S} < q_S \gzp  \ .
\end{align*}
A similarly simple expression can be found to assess whether or not the dark Higgs boson is
lighter than the splitting between the two fermion states $\chi_1$ and $\chi_2$. When $\mdone <0$, one has
$M_S < \dchi$ when
\begin{align}
 \lambda_S < 4 (\ysr+\ysl )^2 \ .
\end{align}

While the Higgs quartic and the dark gauge coupling are a priori free parameters of our model, we can place an upper bound by requiring them to remain perturbative at least up to the TeV scale.\footnote{Assumed to be the typical mass scale of the SM-charged vector-like fermions creating the kinetic mixing parameter.} Considering the gauge coupling alone leads to the usual result $\azp \equiv \gzp^2/ (4\pi) \lesssim 0.5$. On the other hand, and especially in the pseudo-Dirac regime where $\ysl,\ysr \ll 1$, the Higgs quartic beta function is given by
\begin{align}
	\beta_{\lambda_S} = 96 \gzp^2 + 20 \lambda_S^2 - 48 \gzp^2 \lambda_S \ ,
\end{align}
and is positive. In the naive assumption of constant gauge coupling we get $\lambda_S^{1 \textrm{TeV}} =  \lambda_S^{100 \textrm{MeV}} \exp (8  \left( \frac{\azp}{0.1}\right)^2)$. In practice, perturbativity of the dark Higgs quartic hence typically limits $\azp\lesssim 0.1 - 0.15$ depending on the size of the negative contributions from the Yukawa coupling $\ysl, \ysr$ and of the initial value of $\lambda_S$. In the rest of this paper, we make sure that our parameter space satisfies
perturbativity by evaluating the spectrum obtained by using
SPheno~\cite{Porod:2003um,Porod:2011nf} code created by 
\textrm{SARAH}~\cite{staub_sarah_2008,Staub:2012pb,Staub:2013tta}. In particular, we run
the initial couplings up to the TeV scale to check that they remain perturbative, then
evaluate all masses at the sub-GeV scale we are interested at tree-level.

\subsection{Simplified models and dark Higgs boson}\label{sec:dm}

Our numerical results presented in the rest of this Section are based on a scan of the
parameter space of our model (see Appendix~\ref{sec:parameters} for the input parameters and the
chosen range) to identify all relevant DM regions.  We used
\texttt{MultiNest}~\cite{Feroz:2008xx} to direct the scanning process toward relic density values compatible with the result from
the Planck Collaboration~\cite{Ade:2015xua} $\Omega h^2 = 0.1188 \pm 0.0010$. The code BayesFITS is used to interface all the public codes used, including a
slightly modified version of MicrOMEGAs~v.4.3.5~\cite{Belanger:2014vza}, SPheno, and a heavily
modified version of the code BdNMC from~\cite{deNiverville:2016rqh}.

\begin{figure}[t]
	\centering
	\subfloat[]{%
		%\label{fig:a}%
		\includegraphics[width=0.3\textwidth]{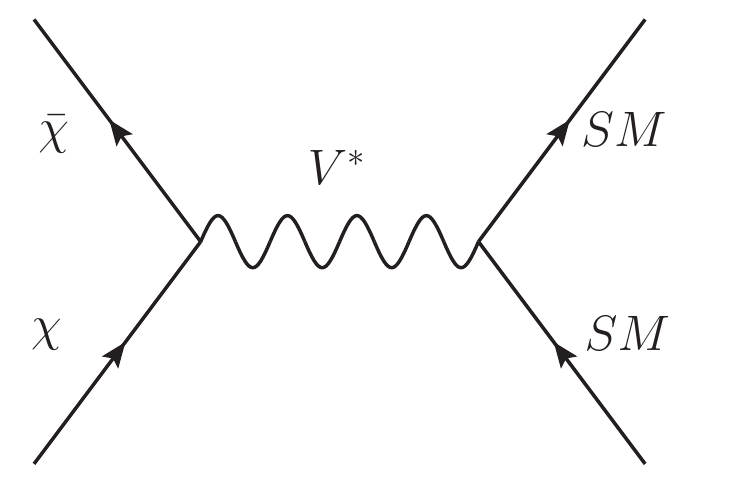}
	}%
	\hspace{0.02\textwidth}
	\subfloat[]{%
		%\label{fig:a}%
		\includegraphics[width=0.3\textwidth]{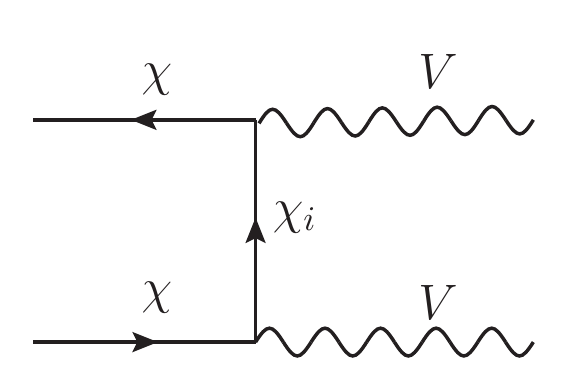}
	}%
	\hspace{0.02\textwidth}
\subfloat[]{%
	%\label{fig:a}%
	\includegraphics[width=0.3\textwidth]{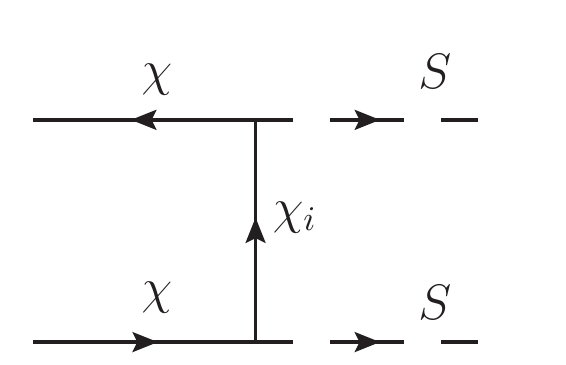}
}%
	\caption{Dominant DM annihilation channels (note that $\chi$ can be either $\chi_1$ or $\chi_2$ depending whether or not coannihilation is relevant). The processes are (a)  S-channel annihilation through an off-shell dark photon, (b) Thermally-induced forbidden annihilation into dark photon, (c) Secluded annihilation into two dark Higgs boson.}
	\label{fig:RelicDensity}
\end{figure}

In this simple model, the correct relic density of DM is typically obtained through
three main diagrams shown in Figure~\ref{fig:RelicDensity}. Crucially, each of them
 suppresses the annihilation of DM at the time of CMB recombination, as
required by the stringent bounds on the annihilation process~\cite{Slatyer:2015jla}. In particular,
the $s$-channel annihilation corresponding to the diagram of Figure~\ref{fig:RelicDensity}(a) is
either population-suppressed in the pseudo-Dirac case $\chi_1 \chi_2 \rightarrow V^*
\rightarrow e^+ e^-$ since all $\chi_2$ states have decayed at recombination time, or it
is velocity-suppressed in the Majorana DM case since $\chi_1 \chi_1 \rightarrow V^*
\rightarrow e^+ e^-$ is a $p$-wave process. Next, while the $t$-channel annihilation of
Figure~\ref{fig:RelicDensity}(b) into dark photons is an $s$-wave process, it has to
proceed through thermal effects in the kinematically forbidden regime $\mdm < \mzp $ to
give the correct relic density. Finally, the $t$-channel annihilation into dark Higgs
bosons of Figure~\ref{fig:RelicDensity}(c) is also of $p$-wave type and thus velocity-suppressed.
Note that the annihilation into $SV$ can also lead to the correct relic
density in a similar fashion as the $VV$ channel, albeit in a smaller region of the
parameter space (we include this region into the larger ``forbidden regime'' one in the following).
%%%%%%%%%%%%%%%%%%%%%%%%%%%%%%%%
%%%%%%%%%%%%%%%%%%%%%%%%%%%%%%%%

%\LD{Add small discussion about the mixed $SV$ final states}

Based on these processes, we have identified four typical scenarios where one can obtain the
correct DM relic density while avoiding CMB bounds: the inelastic DM regime (iDM), the
Majorana DM regime (mDM), the secluded regime and the forbidden regime. These four scenarios
are represented in the $\ms/\mdm$ plane in Figure~\ref{fig:DMregions}.  As was underlined
in~\cite{Darme:2017glc}, when the dark Higgs boson is long-lived (typically $\ms < 2\mdm$
and $ \ms < \mzp$), the relic density must be obtained by considering a two-component
DM-like scenario to obtain simultaneously the correct relic density and the dark Higgs boson
metastable density. Let us review the properties of each scenario in turn:

\begin{figure}[t]
	\centering
	\subfloat[]{%
		%\label{fig:a}%
		\includegraphics[width=0.45\textwidth]{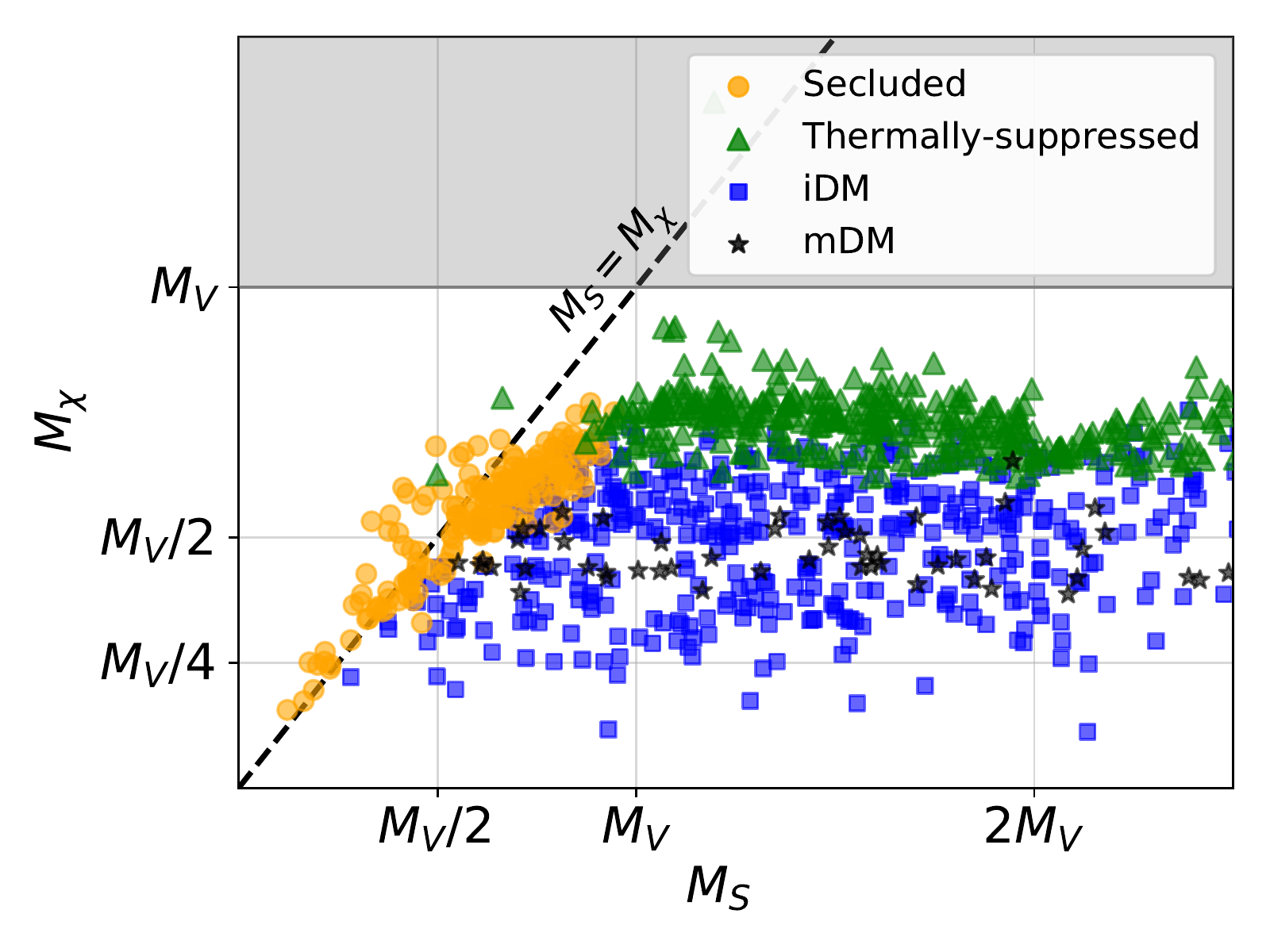}
	}%
	\hspace{0.02\textwidth}
	\subfloat[]{%
		%\label{fig:a}%
		\includegraphics[width=0.45\textwidth]{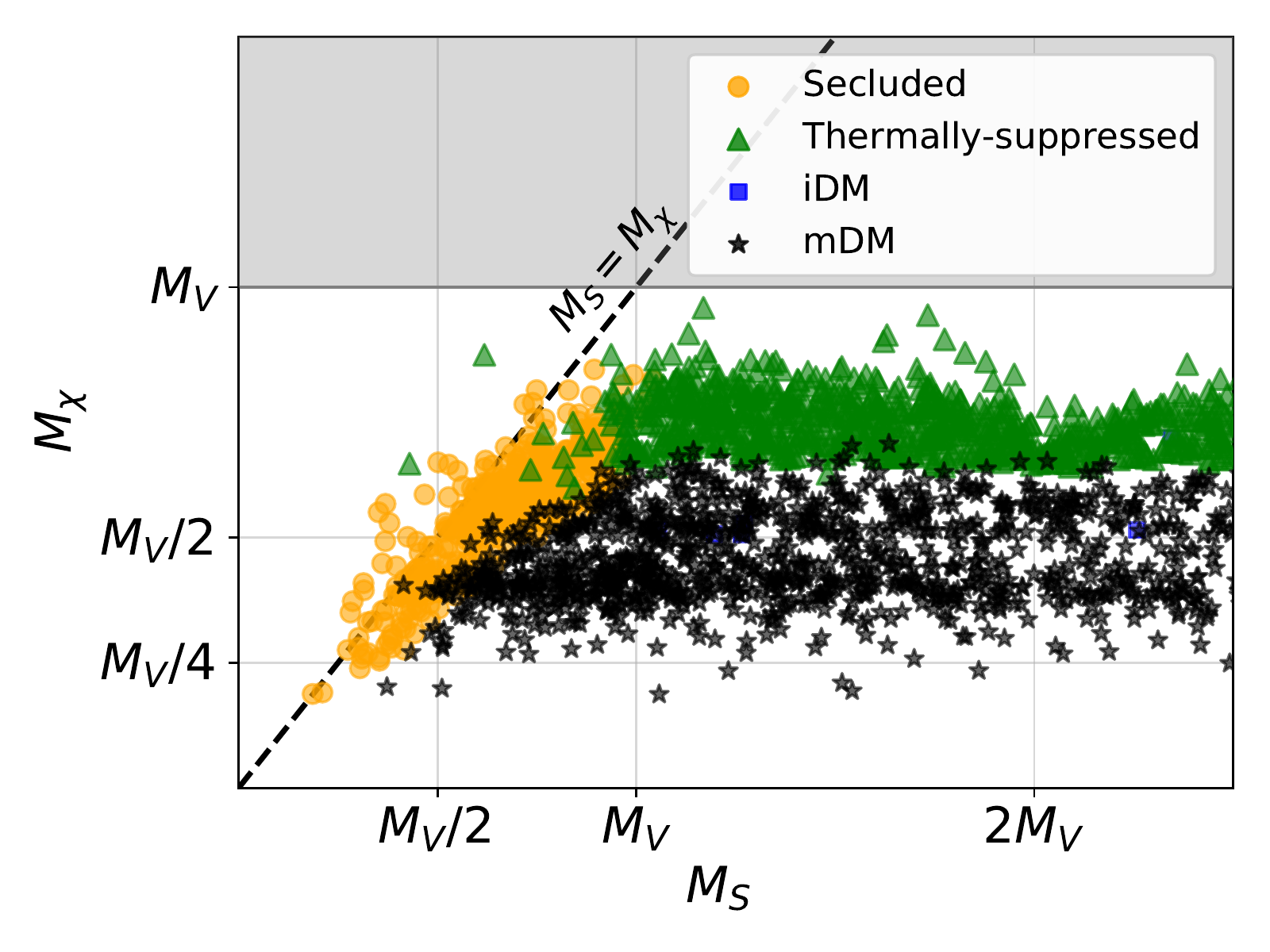}
	}%
	\caption{Data points satisfying BBN bounds and the relic density constraint from
	the numerical scan described in the main text, represented in the $(\ms,\,|\mdm|)$
plane with masses in unit of $\mzp$ for (a) $\dchi < 0.5 \mdm$ and (b) $\dchi > 0.5 \mdm$. We
have classified them in four regimes: inelastic DM (blue squares), Majorana DM (black
stars), secluded regime (orange circles), and forbidden regime (green triangles).}
	\label{fig:DMregions}
\end{figure}

% 
% 
% The dark Higgs boson with sizable Yukawa couplings modifies this picture in two different ways: by providing a supplementary annihilation $t$-channel when it is kinematically allowed, by enhancing the normal $s$-channel annihilation and finally b

\begin{itemize}
 \item \textit{Inelastic DM regime (iDM)}: This scenario has been the most studied one in
	 the literature. It roughly corresponds to the mass spectrum $\mdm < \ms,\mzp$
	 with $\mdone <0$ in our convention and $\dchi < 0.5 \mdm$ as can be seen in
	 Figure~\ref{fig:DMregions}.  In this regime the 
 role of the dark Higgs boson is typically neglected and an approximate number symmetry is
 introduced so that $\ysr,\ysl \ll 1$. Under this assumption, DM is of the pseudo-Dirac
 type and  the  mechanism to obtain the correct relic density is a coannihilation through
 an off-shell dark photon. This scenario has the advantage of achieving the correct
 relic density for a wide range of parameters and leading to strong signatures in
 accelerator experiments. A typical spectrum in this scenario is
\begin{align}
  \mzp : \mdm : \dchi = 3 : 1 : 0.1 \hskip0.5cm\mathrm{and}\hskip0.5cm  \ms > \mdm\ .
\end{align}
with couplings of typical order	$g_{11} \sim 0$, $g_{12} \sim \gzp$ and $\ydm \ll 1$.
Note that a priori the dark Higgs boson mass is a free parameter in this scenario, as long
as $\mdm < \ms $. We show in Figure~\ref{fig:metastab} the transition
between the two regimes : $\mdm < \ms $ and $\mdm > \ms $ (secluded regime, described
below). An interesting observation is that one should strictly speaking also impose $\mhds
< \ms $ since for $ \mdm < \ms < \mhds$ the dark Higgs boson already modifies
significantly the relic density by depleting the abundance of the heavy state $\chi_2$
and hence reducing the efficiency of the coannihilation mechanism. For larger values of
$\dchi$ (above $0.1$), this scenario is in fact already quite constrained by
accelerator-based experiments if the $\chi_1$ is assumed to make for the whole DM relic
density, as we will see in the next sections.
\begin{figure}[t]
	\centering
%	\subfloat[]{%
		%\label{fig:a}%
		\includegraphics[width=0.6\textwidth]{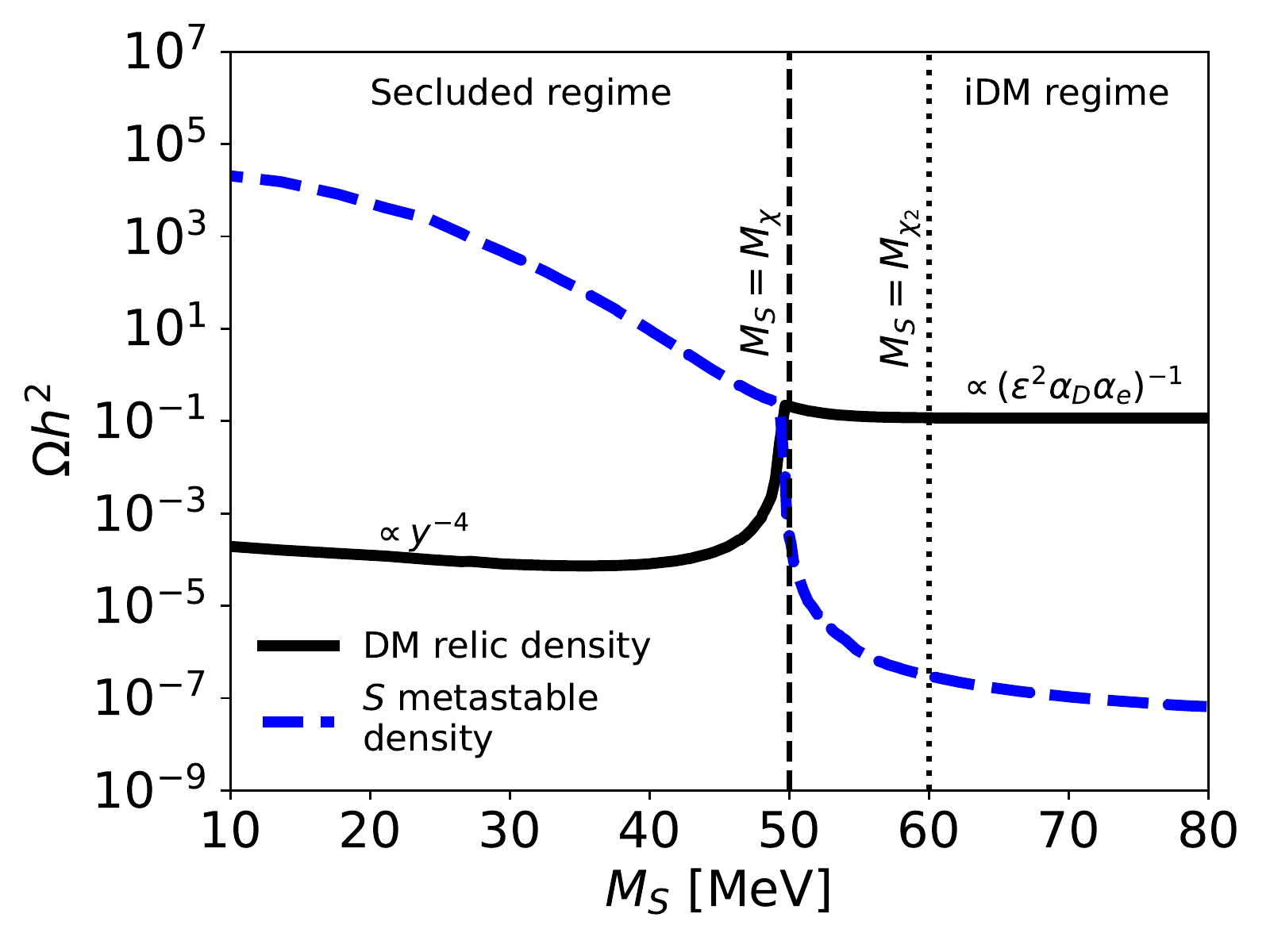}
%	}%
	\caption{Dark matter relic density $\Omega h^2$ and dark Higgs boson metastable
	abundance  $\Omega h^2_S$ (normalized to the relic density it would have had today if
it had been stable) as function of the dark Higgs boson mass. We have chosen $\mdm = 50$ MeV,
$\mhds = 60$ MeV, and the rest of the couplings so that the correct relic density is
obtained in the iDM regime when $\ms > \mhds$.}
			\label{fig:metastab}
\end{figure}
 \item \textit{Majorana regime (mDM)}: When the constraints on the Yukawa couplings are
	 relaxed and $\ysr, ~\ysl \sim 1$, the Majorana mass can be of the same
	 order or larger as the Dirac mass, leading to a Majorana DM candidate with
	 sizable coupling both to the dark Higgs boson and to the dark photon. The mass
	 spectrum in this regime is typically $\mdm < \ms,\mzp$, with $\mdone >0$ in our
	 convention and $\dchi \sim \mdm$ as can be seen in Figure~\ref{fig:DMregions}.
% The mass of the heavy Majorana state $\chi_2$ is then only mildly related to the one of
 %the DM $\chi_1$, so that large splitting between these two states typically occurs. 
 This
 scenario shares many similarities with the iDM one but can lead to different limits at
 accelerator-based experiments. A typical mass ratio in this scenario is
\begin{align}
  \mzp : \mdm : \dchi = 3 : 1 : 1 \ ,
\end{align}
where $\mdm$ is taken positive.\footnote{As described in the Appendix~\ref{sec:parameters}, we then have to fix $\ysl$ and $\ysr$ so as to be consistent with this mass ratio.}
Like in the previous case, as long as $\mdm < \ms \ll \mzp$ the dark Higgs boson mass
usually does not play a role in the determination of the relic density. In addition, and similarly to the iDM regime, the parameter space where $\chi_1$ is the dominant component of DM is already severely constrained by accelerator searches. %Notice, the fact that we do not
%rely on coannihilation means that the DM annihilation is typically more effective, leading
%to the correct relic density slightly lower values of $\eps$ than in the iDM case.
 \item \textit{Secluded regime}: When the dark Higgs boson mass becomes of the same order or lighter than the
	 DM particle $\ms \sim \mdm$, the relic density is mainly fixed by $t$-channel annihilation into a pair
	 of dark Higgs bosons (Figure~\ref{fig:RelicDensity}(c)), corresponding to the
	 left-hand side of Figure~\ref{fig:metastab}. Since the annihilation proceeds
	 through an unsuppressed $t$-channel process, low values $\ysr,\ysl \ll 1$ are
	 typically preferred to obtain the correct relic density, which in turn leads to
	 small splitting between $\chi_1$ and $\chi_2$. Notice that our ``secluded' regime differs slightly from the one of~\cite{Battaglieri:2017aum} since we consider annihilation into the scalar dark Higgs boson and not into the dark photon. Accelerator signatures of this
	 regime are therefore similar to the iDM regime, but the relic density is
	 independent of the kinetic mixing, leading to a larger range of accessible
	 parameter space. A typical mass ratio for this scenario is
\begin{align}
  \ms : \mdm : \mzp : \dchi = 5/6 : 1 : 5/2 : 0.1 \ .
\end{align}
Note that in general this scenario can simultaneously be probed by accelerator-based experiments and
BBN/CMB observables since the dark Higgs boson can have a very large metastable density after
freeze-out, as shown in Figure~\ref{fig:metastab}. In the regime where $\ms  \gtrsim
\mdm$ and $\ms \lesssim \mzp$ the DM relic
density is fixed by a mechanism sharing similarities with the sterile coannihilation presented
in~\cite{DAgnolo:2018wcn}. Overall, the relic density of the whole dark sector (including dark Higgs
boson and DM) depends on the thermally-suppressed process $ S S \rightarrow V V$
and hence on the dark photon mass.\footnote{While typical metastable dark Higgs boson abundance our of reach of BBN bounds, they are nonetheless constrained by CMB bounds on decaying DM if the dark Higgs boson is extremely long-lived.}
  \item \textit{Forbidden regime}: When the DM mass moves closer to the dark photon one,
	  the $t$-channel annihilation into dark photons of
	  Figure~\ref{fig:RelicDensity}(c) becomes dominant due to thermal effects, this is
	  the so-called ``forbidden regime''. In this setup, our model resembled the one
	  of~\cite{DAgnolo:2015ujb}, albeit without a pure Dirac DM scenario. A typical
	  mass ratio in this scenario
	  is
\begin{align}
  \ms : \mdm : \mzp = 6/5 : 4/5 : 1 \ .
\end{align}
Similarly to the previous case, the relic density is then independent of the kinetic mixing
parameter. On the other hand, the dark photon decays visibly leading to many more
signatures at accelerator experiments. Both the heavy state $\chi_2$ and the dark Higgs
boson have little impact on the relic density. An interesting exception occurs in the 
% isit really left-right symmetry in the usual sense of the term?
limit $\ysr \simeq \ysl $, for which the coupling between the Majorana DM and the gauge
boson approximately vanishes. In this limit, an annihilation process of the form $\chi \chi
\rightarrow V V$ can nonetheless proceed through an $s$-channel dark Higgs boson exchange,
albeit at a slightly suppressed rate, which in turn leads to DM masses closer to the dark
photon one. Notice that for really small gauge coupling $\gzp \sim 0.01$, the relic t-channel annihilation is suppressed enough to lead to the correct relic density for $\mdm \sim \mzp$, but is excluded by CMB bounds. 
\end{itemize}

%\section{Sommerfeld effects for light dark Higgs}

\subsection{Astrophysical and cosmological constraints on the dark Higgs boson}

While the dark Higgs boson is an important ingredient of the scenarios described above, its long lifetime when $\ms < \mzp$ and $\ms < 2 \mdm$ can
also lead to various cosmological constraints.

\paragraph*{BBN and other cosmological bounds.}

%Here, we consider the dark Higgs mass to be below the dimuon threshold and in which case
%the only decay channel available is $e^+e^-$. 

The effect of long-lived scalars like the dark Higgs boson on BBN was studied in
\cite{Fradette:2017sdd} and it was seen that stringent constraints on $\eps$ can be
placed, restricting the lifetime of a scalar like dark Higgs boson to $<0.1$ s.  However, it
was shown in \cite{Darme:2017glc} that such strong constraints can be avoided with the
addition of a DM candidate as considered here.  In particular, constraints from
early time energy injection ($t\lesssim 10$ s) can be avoided through the efficient
annihilation channels (for e.g.. $SS\to\chi\chi$) that are possible in this model, while
late time energy injection ($t\gtrsim 100$ s) constraints are mitigated by kinematically
restricting the decay modes of dark Higgs boson to leptonic products only.

For longer lifetimes (corresponding to $\eps \lesssim 10^{-6}$), bounds from CMB
deformation (see, e.g.~\cite{Slatyer:2016qyl}) due to dark Higgs boson decays during the
recombination era becomes relevant and can probe a dark Higgs boson metastable density up
to $\Omega h^2_S \sim 10^{-12}$. 
%This mainly constraints point 

Furthermore, when the DM mass is around $5$ MeV and below, CMB bounds on the effective number of neutrinos arise~\cite{Nollett:2013pwa}, with some model dependency on the precise value of the limit. In the same mass range and for strong gauge coupling $\alpha_D \gtrsim 0.5$, constraints from self-interaction of the DM may also become relevant~\cite{Battaglieri:2017aum}. 
%Given the model dependence of these two types of bounds  and the fact that we have practically no points in the relevant regions,  ranges are  , \LD{Complete}

\paragraph*{ Constraints from supernova 1987A.}

Constraints on light dark sector particles in the MeV range from supernova 1987A (SN1987A)
\cite{Burrows:1986me, Burrows:1987zz, Raffelt:1987yt, Raffelt:1996wa} have been studied in
several articles \cite{Bjorken:2009mm, Dent:2012mx, An:2013yfc, Redondo:2013lna,
Kazanas:2014mca, Rrapaj:2015wgs, Chang:2016ntp, Hardy:2016kme, Mahoney:2017jqk,
Chang:2018rso}.  We briefly comment about the relevance of the constraints obtained in
\cite{Chang:2018rso} for an inelastic DM model that is similar to the model studied here.

A core-collapsed supernova like SN1987A gives rise to a hot neutron star or a
proto-neutron star environment. The creation of new particles that can interact with SM
particles and their subsequent escape can lead to cooling of the interior of the
proto-neutron star.  The core of the proto-neutron star is around $T\sim 30$ MeV and a
significant loss of energy in the form of dark sector particles in the MeV range can
result in the supernova cooling at a faster rate than expected.  In previous studies, the
dark photon production inside the supernova core through bremsstrahlung process has been
used to constrain the kinetic mixing parameter $\eps$.  A recent update of this constraint
includes a DM candidate in addition to the dark photon \cite{Chang:2018rso}.  In the
context of the model studied here, the dark sector includes a further addition of the dark Higgs boson. 

%As discussed in ref.~\cite{Chang:2018rso}, the upper boundary of the constraint on kinetic
%mixing parameter is obtained by defining a thermal decoupling radius, inside which the
%dark sector particle is in thermal equilibrium and outside which it is free streaming.
%This decoupling radius is evaluated by equating the observed neutrino luminosity with the
%blackbody luminosity of the dark sector particles from a radius $R_d$, which is the
%decoupling radius.  A further approximation is that the decoupling radius is determined by
%solving only for kinetic equilibrium but not chemical, since it is expected that the
%chemical decoupling radius is generally smaller.  Finally, one defines the criteria for
%finding the zone of decoupling as at least one scattering of the DM particle off
%of SM particles (protons), which leads to a constraint on the kinetic mixing parameter
%(see \cite{Chang:2018rso} for more details).

%assuming that the DM particle is trapped inside the supernova and that the DM
%scatters at least once in order for the trapping to occur. It is further assumed that the
%decoupling radius, outside which the dark sector particle is free streaming and inside
%which it remains trapped, is only determined through kinematic processes like scattering.
%In other words, only the kinetic decoupling radius is taken into account while neglecting
%the chemical decoupling radius as it is expected to be smaller than the former (see
%\cite{Chang:2018rso} for more details).

For SN1987A constraints on DM in our model, the inelastic
DM case when $\Delta=0$ is the relevant case considered in \cite{Chang:2018rso}.  Note that the constraints for
$\Delta>0$ are not directly applicable here, because of the model restrictions used in
\cite{Chang:2018rso}.  In particular, the presence of elastic scattering $\chi_1 p\to
\chi_1 p$ is neglected, unlike the case for our model where it becomes significant when the splitting $\dchi$ increases (see Eq.~\eqref{eq:mDMcoupling}). 
%Considering the range of parameters used in
%the analysis here, this constraint does not have any effect on our results.

%To derive constraints on dark
%Higgs boson, we assume that there is a thermal population of DM trapped inside the
%supernova which can scatter the dark Higgs boson produced inside the supernova core.  The
%$S\chi$ scattering then provides the mechanism for trapping the dark Higgs boson.
%Following the calculation in \cite{Chang:2018rso}, the resulting constraint depends on the
%scattering cross section which in the case of $S\chi$ scattering goes as
%$y_{DM}^2\alpha_D(\ms^2/\mzp)^2$, as opposed to $\eps^2\alpha_D(4\pi\alpha_{em})$ in the
%case of DM particle scattering off protons.  Here, $y_{DM}$ is the coupling between DM and
%dark Higgs boson while $\alpha_D=g_V^2/4\pi$.  The constraint on DM is $\eps\gtrsim
%10^{-6}$, for $\alpha_D=0.1$ and $\mzp=3\mdm$ \cite{Chang:2018rso}, while the constraint
%on dark Higgs boson from $S\chi$ scattering is $y_{DM}\gtrsim 10^{-6}$, for
%$\alpha_D=0.1,\mzp=3\ms$ MeV and $\mdm=\ms$.  

In principle the effect of this constraint on dark Higgs boson should be mitigated by two reasons: the
production of dark Higgs boson being suppressed compared to dark photon and DM,
and the ambiguity of the distribution of DM and dark photon inside the supernova
particularly for the large values of $\eps$ and other relevant couplings considered here,
for which the dark Higgs boson can scatter off dark photon and/or DM at a
significantly large rate.  Obtaining a proper lower
limit on the parameter space will require a calculation involving full simulation of the
dark sector population inside the supernova.

\section{Dark Higgs boson at accelerator experiments}\label{sec:bdump}

We present in this section the bounds on the four scenarios presented above that arise from
accelerator-based experiments. The fact that the dark sector in our model not only
contains
a dark photon and a DM particle, but also a dark Higgs boson and a heavy dark sector
state $\chi_2$ will significantly enhance the prospect of its experimental verification.

Most generic searches typically focus on a dark photon and either assume that it
decays invisibly and search for missing-energy
signatures~\cite{Andreas:2013lya,Izaguirre:2014bca} - the strongest bounds using this
strategy are currently from BaBar~\cite{Lees:2017lec} and NA64~\cite{Banerjee:2017hhz}.
Alternatively, one can also try to observe its decay as bumps in a dilepton invariant
spectrum such as in NA-48/2~\cite{Batley:2015lha}, BaBar~\cite{Lees:2014xha} and
LHCb~\cite{Aaij:2017rft}. If the dark photon decays visibly but with a longer lifetime
then a range of long-baseline experiments (see~\cite{Alexander:2016aln} for a review)
further probe the range $\eps \lesssim 10^{-4} - 10^{-5}$. 

One can also search for scattering of DM particles produced through on-shell dark
photon decay in various beam dump experiment.  This has been an active field of research
for a
decade~\cite{Batell:2009di,Batell:2014mga,Izaguirre:2015yja,deNiverville:2016rqh,Battaglieri:2016ggd,Izaguirre:2017bqb,Aguilar-Arevalo:2018wea},
and we have included the case of scattering in the LSND
experiment as an example of the current sensitivity.\footnote{While scattering bounds also exist for E137 and recently for
	miniBooNE, they have typically a reach relatively similar to
	LSND~\cite{Aguilar-Arevalo:2018wea}. We therefore focus on the LSND case in the
following.}

In this work, while including the previous bounds, we focus instead on stronger but
more model-dependent bounds arising from the decay of either the heavy dark sector state
$\chi_2$ or of the dark Higgs boson. In particular, we complement our previous
work~\cite{Darme:2017glc,Darme:2018pop} in several directions by including more production
channels for the dark Higgs boson and by considering also its decay channel through an
off-shell dark photon, which becomes dominant when $\ms > \mzp$. Furthermore, we have also
included the bounds from the decay of the long-lived heavy dark sector state $\chi_2
\rightarrow \chi_1 e^+ e^-$, studied in~\cite{morrissey} in a supersymmetric context and
later by~\cite{Izaguirre:2017bqb} in a setup similar to our iDM regime. Being obtained
with a completely different code, our results can be considered as an independent cross-check for
the iDM regime.

\subsection{Dark Higgs boson production and decay}

\begin{figure}[t]
	\centering
	\subfloat[]{%
		%\label{fig:a}%
		\includegraphics[width=0.45\textwidth]{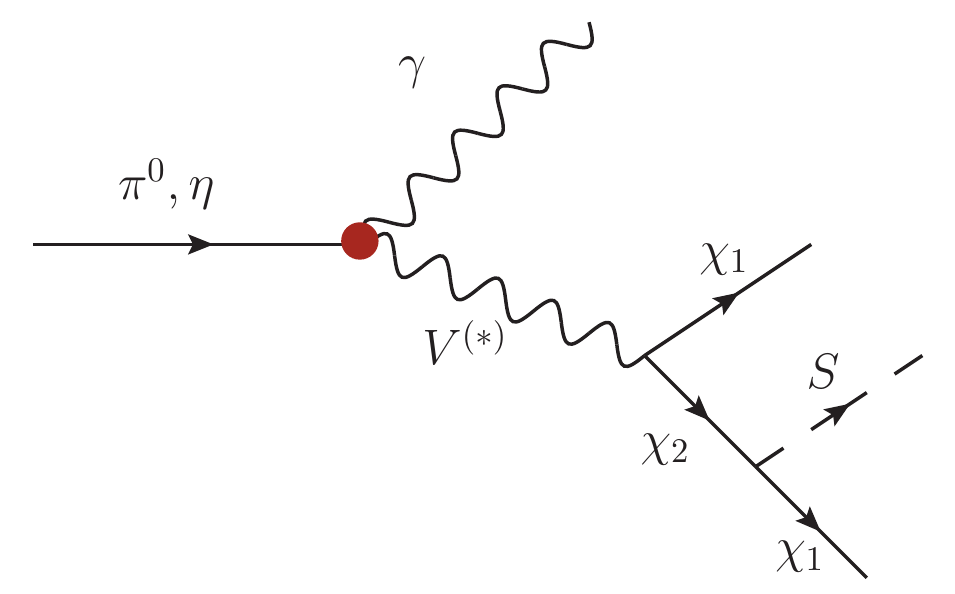}
	}%
	\hspace{0.02\textwidth}
	\subfloat[]{%
		%\label{fig:a}%
		\includegraphics[width=0.45\textwidth]{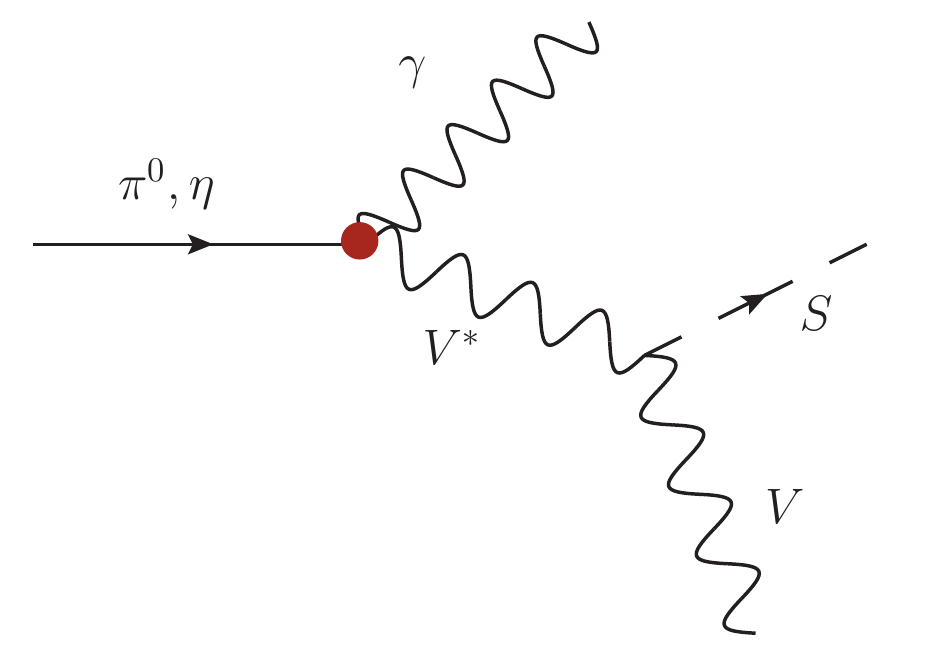}
	}%
	\hspace{0.02\textwidth}
	\caption{ Dark Higgs boson production in meson decay through (a) chain decay of a
	heavy dark sector state and (b)  dark Higgstrahlung.}
	\label{fig:DHprod}
\end{figure}

In accelerator-based experiment, the dark Higgs boson is mainly produced through either
the chain decay of some heavy dark sector states (for example $\chi_2$, as shown in
Figure~\ref{fig:DHprod}(a)), when this is kinematically allowed, or through dark
Higgstrahlung after an excited dark photon is produced as shown in
Figure~\ref{fig:DHprod}(b). In general, the former has higher rates but at the expense of
a stronger model dependence. Both processes can happen either in proton beam dump
experiments, for instance from meson decay, or in electron beam-dump experiment through
bremsstrahlung production of dark photon (dark bremsstrahlung). 

For the  dark bremsstrahlung production (in our case relevant for the electron beam dump
experiment E137) we have used the public code Madgraph~\cite{Alwall:2014hca}.  Some details
about numerical setup and the target form factors used in the calculation are given in Appendix B.
%\LD{Add all references and some details about the procedure here}.

For the meson decay cases, we use a modified version of the code
BdNMC~\cite{deNiverville:2016rqh} where we have directly included the expression for the various
differential branching ratios relevant to both the dark Higgs boson and heavy dark sector
states production. We refer to~\cite{Darme:2017glc} for the details of the dark
Higgstrahlung process in the case of proton beam dumps. In this work, we will focus on the
remaining expressions which are needed for the dark sector state production.

In order to treat the chain decay case in proton beam dump, we have recalculated the
production rates for DM in meson decays through an on-shell or off-shell dark photon
including the proper rotation matrices for the dark sector states $\chi_1$ and $\chi_2$.
This allows one to smoothly interpolate between the various regimes described above (in
particular, we recover the standard results from ref.~\cite{Izaguirre:2017bqb} for the iDM
regime). As in~\cite{Darme:2017glc}, we are interested in the differential decay rate
$d \text{BR}_{\piz \rightarrow \gamma \chi_i \chi_j}/ds d
\theta^V $, where $\theta^V$ denotes the angle between $\chi_i$ and the mediating dark photon $V^*$
in the rest frame of the latter and $s$ denotes the four-momentum squared of $V^*$. A key
parameter is the coupling $g_{ij}$ between $\chi_i \chi_j$ and $V$, which is already described
analytically in Eq.~\eqref{eq:iDMcoupling} and Eq.~\eqref{eq:mDMcoupling}. In term of the
rotation matrix $Z^X$ relating the DM gauge eigenstates to the mass eigenstates, it is
given by
\begin{align}
 g_{ij} = \gzp (Z^X_{i1} Z^X_{j1} - Z^X_{i2} Z^X_{j2}) \ ,
\end{align}
and the vertex includes a $\gamma^\mu \gamma^5$ factor in our conventions (real mixing
matrices but potentially negative mass for $\chi_1$). We obtain
\begin{align}
  \frac{d^2 \text{BR}_{\pi^0 \rightarrow \gamma \chi_i \chi_j}}{ds d \theta^V} = \mathcal{S} & \times \text{BR}_{\pi^0 \rightarrow \gamma \gamma} \times g_{ij}^2 \times \nn \\
  &  \frac{\eps^2 \azp}{4\pi}  \ s \left(  1 - \frac{s}{m^2_{\piz}}\right)^3  \times \frac{ \sqrt{\lambda}~ \left(2 s [s-(M_i+M_j)^2] - \lambda \sin^2 \theta \right)}{(s-\mzp^2)^2 + \mzp^2 \Gamma^2_V}\sin \theta \ ,
  \label{eq:diffrate}
\end{align}
where $\mathcal{S}$ is a symmetry factor equal to $1/2$ if $i = j$ and $1$ otherwise, $\Gamma_V$ is the width of the dark photon and the triangular function $\lambda$ is defined as
\begin{align*}
 \lambda ~\equiv~ \left(1 - \frac{(M_i+M_j)^2}{s}\right)\left(1 - \frac{(M_i-M_j)^2}{s}\right) \ .
\end{align*}
As usual, this straightforwardly applies to the case of the $\eta$ meson by replacing $m_{\pi^0}$ by $m_{\eta}$ and $\text{BR}_{\piz \rightarrow \gamma \gamma}$ by $ \text{BR}_{\eta \rightarrow \gamma \gamma} = 0.394$. The subsequent decay of the heavy dark sector state $\chi_2 = \chi_1 S$ is then considered
to occur instantaneously and with a branching ratio of one if it is kinematically
accessible.

%%%%%%%%%%%%%%%%%%%%%%%%%%%%%%%%
%%%%%%%%%%%%%%%%%%%%%%%%%%%%%%%%

We are interested in signatures corresponding to the decay product of the dark Higgs boson. In the mass range we consider and if $ \ms < 2
\mdm, \ms < \mzp$, the dominant decay channel is a loop-induced decay into an
electron-positron pair, leading to an extremely long
lifetime~\cite{Batell:2009yf,Darme:2017glc},
\begin{align}
\label{eq:Slifetime}
\tau_S \propto 10 \textrm{ s} \times \left( \frac{\aem}{q_S^2  \azp}\right) \left(  \frac{ 10^{-3}}{\eps}\right)^4 \left( \frac{50 \textrm{ MeV}}{\ms}\right) \left( \frac{\mzp}{ 100 \textrm{ MeV}}\right)^2 \ .
\end{align}

On the other  hand if $\mzp + 2 m_e < \ms < 2 \mdm$, the decay channel $S \rightarrow V
e^+ e^- $ becomes kinematically accessible. Interestingly, for relatively small splitting,
the decay width takes a simple form
\begin{align}
	\Gamma_S = \frac{16 q_S^2 \azp \aem \eps^2}{15 \pi} \times\frac{(\ms -
	\mzp)^5}{\mzp^4},
\end{align}
which can be parametrically expressed as 
\begin{align}
\label{eq:Slifetime_short}
c \tau_{S}  \propto 20 \text{ m } \times  \left(  \frac{0.1}{\alpha_D} \right)  \left(  \frac{ 10^{-4}}{\eps}\right )^2 \left( \frac{0.1 \mzp }{\ms - \mzp}\right)^5  \left( \frac{100 \textrm{ MeV}}{ \mzp } \right) \ .
\end{align}
As it was already noticed in~\cite{Essig:2009nc} in a related context, the dark Higgs
boson lifetime is now significantly shorter and scales only as $\eps^2$, similarly to the
heavy dark sector decay considered in~\cite{Izaguirre:2017bqb}. This allows one to probe
extremely small values of $\varepsilon$ since a sizable fraction of the dark Higgs bosons
produced in accelerator-based experiments will decay in the detector.\footnote{It is
	interesting to compare this result with the expression for the lifetime of the
	long-lived heavy dark sector state $\chi_2$ which can be easily derived from the
	results of~\cite{Izaguirre:2017bqb},
\begin{align}
c \tau_{\chi_2}  \propto 100 \text{ m } \times  \left(  \frac{0.1}{\alpha_D} \right)  \left(  \frac{ 10^{-3}}{\eps}\right )^2 \left( \frac{0.2}{\dchi}\right)^5 \left( \frac{25 \textrm{ MeV}}{\mdm}\right)^5 \left( \frac{\mzp}{ 100 \textrm{ MeV}}\right)^4 \ .
\end{align}
}

% 
% In this section, we will focus on
% % LD -- added
% examining the dark Higgs boson detection prospects in 
% % LD -- end
% three proton beam-dump experiments:
% LSND\cite{Athanassopoulos:1996ds}, miniBooNE\cite{AguilarArevalo:2008qa}  and the
% proposed SBND experiment at Fermilab\cite{Antonello:2015lea}. The details of the
% experimental setups are presented in Table~\ref{tab:exppars}. These three experiments rely on proton beams
% with relatively low energy so that we expect dark sector production through
% bremsstrahlung and direct production to be sub-dominant compared to the meson decay
% mechanism\cite{deNiverville:2016rqh}.
% 
% % LD -- Added ``Past``
% Notice that past electron beam-dump experiments, like E137\cite{Riordan:1987aw}, can also lead to dark
% sector beams through dark photon production by bremsstrahlung. However, the bounds on the kinetic
% mixing parameter $\eps$ 
% % LD -- added
% %from the latter
% derived from dark Higgs boson production and decay at these facilities 
% % LD -- end
% were found in\cite{morrissey} (in a context roughly similar
% to ours -- albeit in a supersymmetric model) to be always significantly weaker than the current
% missing energy bound $\eps < 10^{-3}$. The case studied in\cite{Izaguirre:2017bqb},

\subsection{Accelerator experiments constraints}

In the recent years, there has been a surge in interest in examining the potential reach of
upcoming neutrino-related experiments in probing light DM sectors. In this section, we
rather take the approach of considering the currently applicable bounds from the existing
experiments LSND~\cite{Athanassopoulos:1996ds} and E137~\cite{Riordan:1987aw} to explore
the current limits on the four scenarios described above. In particular, we want to point
out in which case the presence of the dark Higgs boson affects strongly the
constraints and phenomenology of the model.  We want to study in detail the signatures
from the secluded, forbidden and Majorana DM regime which have not been studied in details
so far. We refer to the appendix for more details on the Monte-Carlo simulation used to
get the expected number of events.

\begin{figure}[t]
	\subfloat[]{%
		%\label{fig:a}%
		\includegraphics[width=0.45\textwidth]{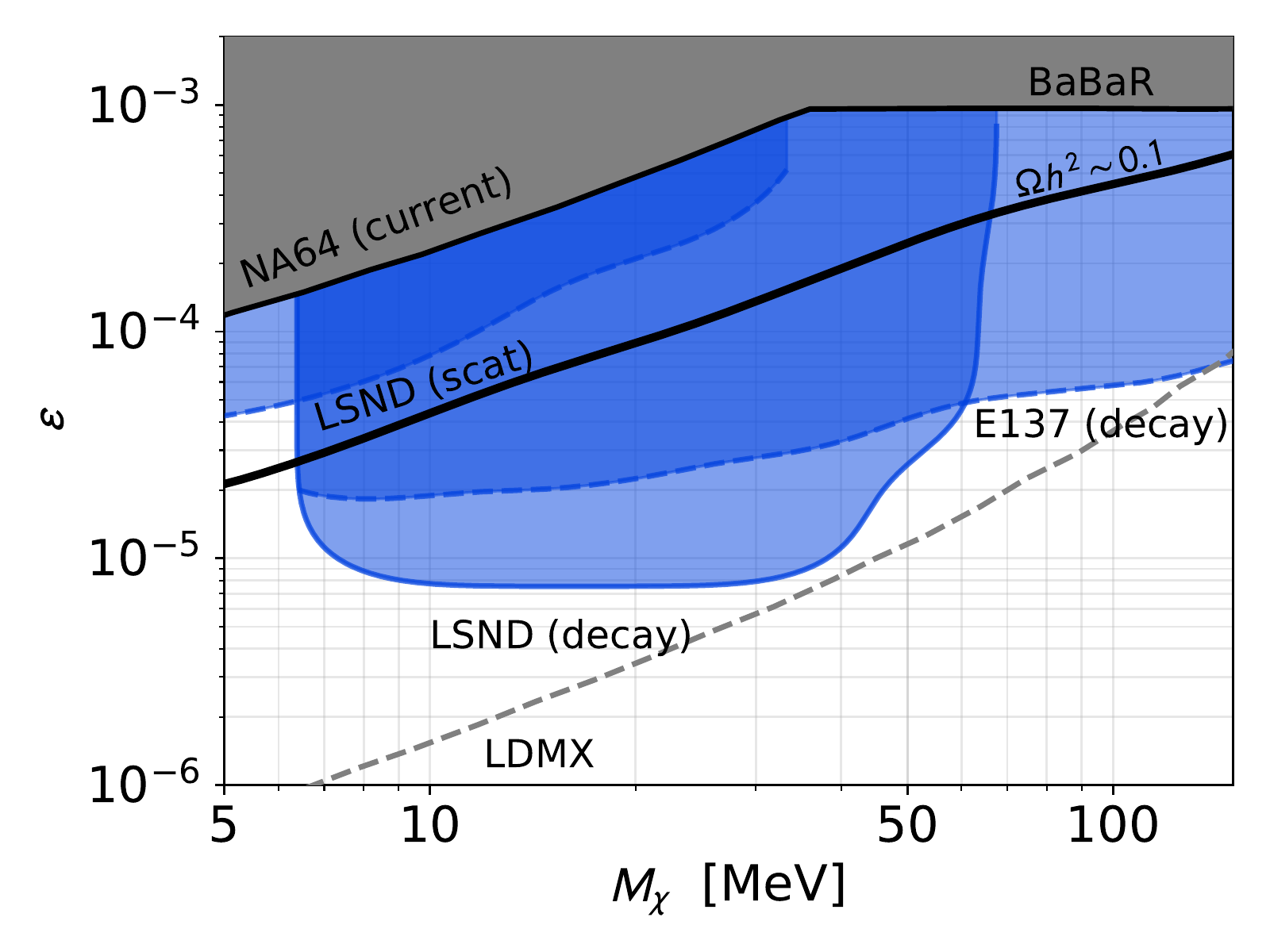}
	}%
	\hspace{0.02\textwidth}
	\subfloat[]{%
		%\label{fig:a}%
		\includegraphics[width=0.45\textwidth]{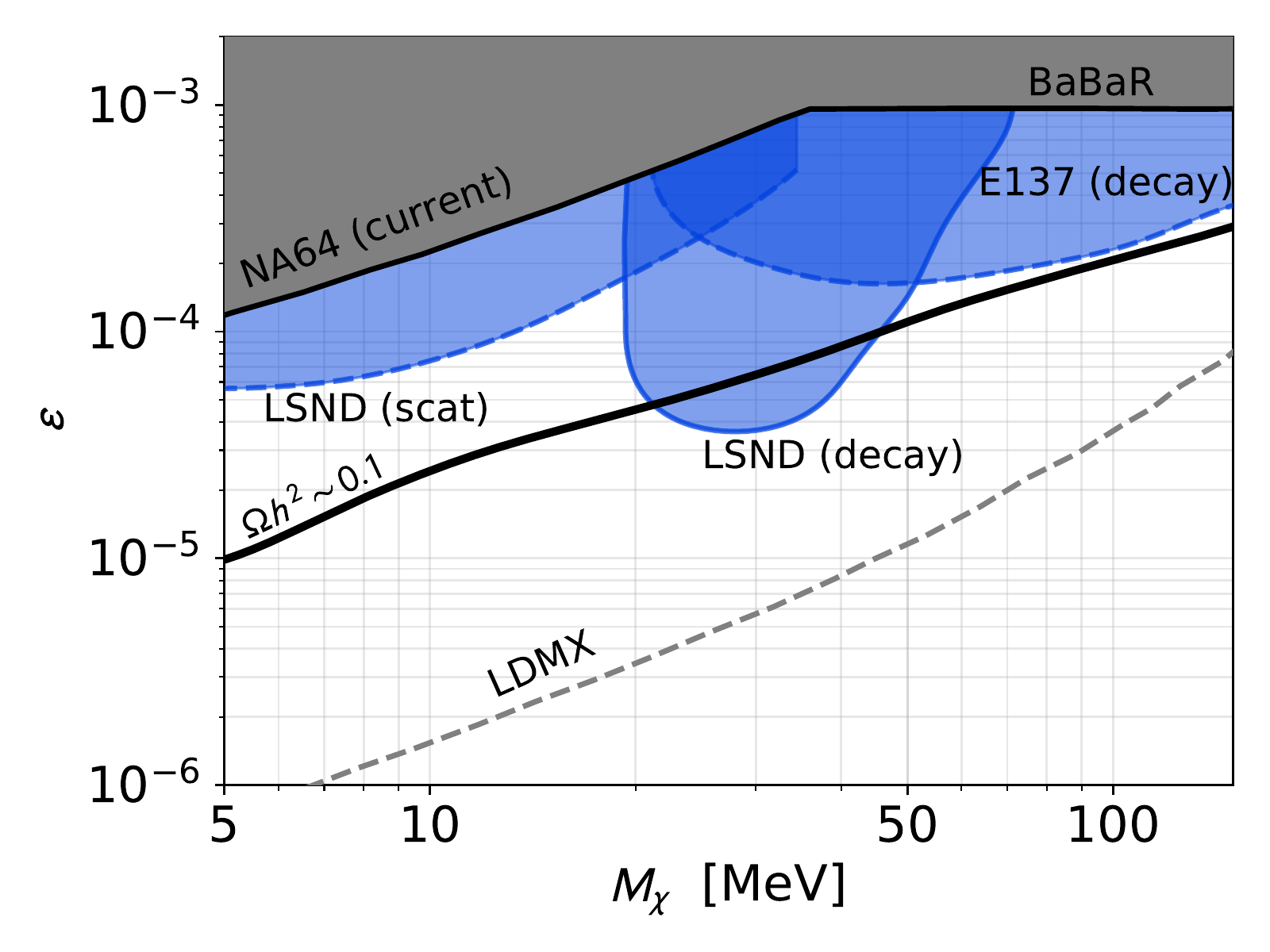}
	}%
	\caption{Bounds on the inelastic DM regime from the E137~\cite{Riordan:1987aw},
	LSND~\cite{Athanassopoulos:1996ds} arising from DM scattering and $\chi_2$ decay
	(blue shaded regions) and missing energy searches (grey regions and grey dashed
	line for the LDMX projection from~\cite{Battaglieri:2017aum}). The thick black
lines correspond to points featuring the correct relic density of DM. We fixed $\azp =
0.1, \ydm = 0.01$ and varied $\mdm$ and $\eps$ using the mass ratios   $\ms : \mdm : \mzp
= 2 : 1 : 3 $ for (a) $\dchi = 0.15 \mdm$ and (b) $\dchi = 0.05 \mdm$.}
	\label{fig:iDMbounds}
\end{figure}

\paragraph{Inelastic DM regime (iDM).} This regime has been already thoroughly analyzed
in the context of both existing experiments and for several prospective ones. Furthermore, as
shown in~\cite{Darme:2017glc}, in this case the dark Higgs boson is extremely long-lived,
so that the bounds from dark Higgs boson-related processes are not relevant. On the other
hand, the decay of long-lived heavy dark sector states give strong signatures in many
existing and upcoming fixed-target and accelerator-based experiments. In order to
cross-check our numerical tools, we have reproduced in Figure~\ref{fig:iDMbounds} the
current constraints which arise from the LSND and E137 experiments and found very good
agreement with the existing literature.

\begin{figure}[t]
	\subfloat[]{%
		%\label{fig:a}%
		\includegraphics[width=0.45\textwidth]{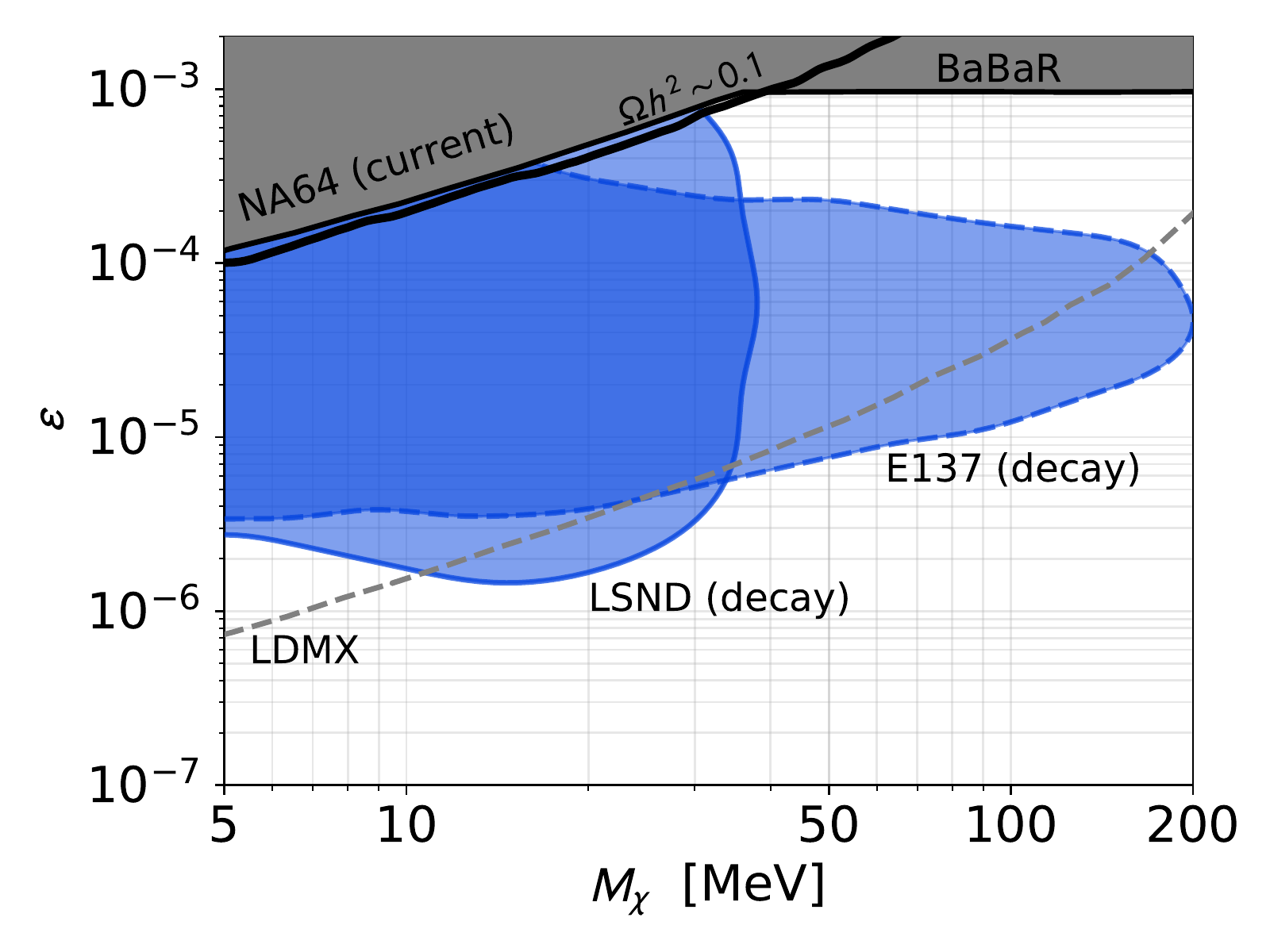}
	}%
	\subfloat[]{%
	%\label{fig:a}%
	\includegraphics[width=0.45\textwidth]{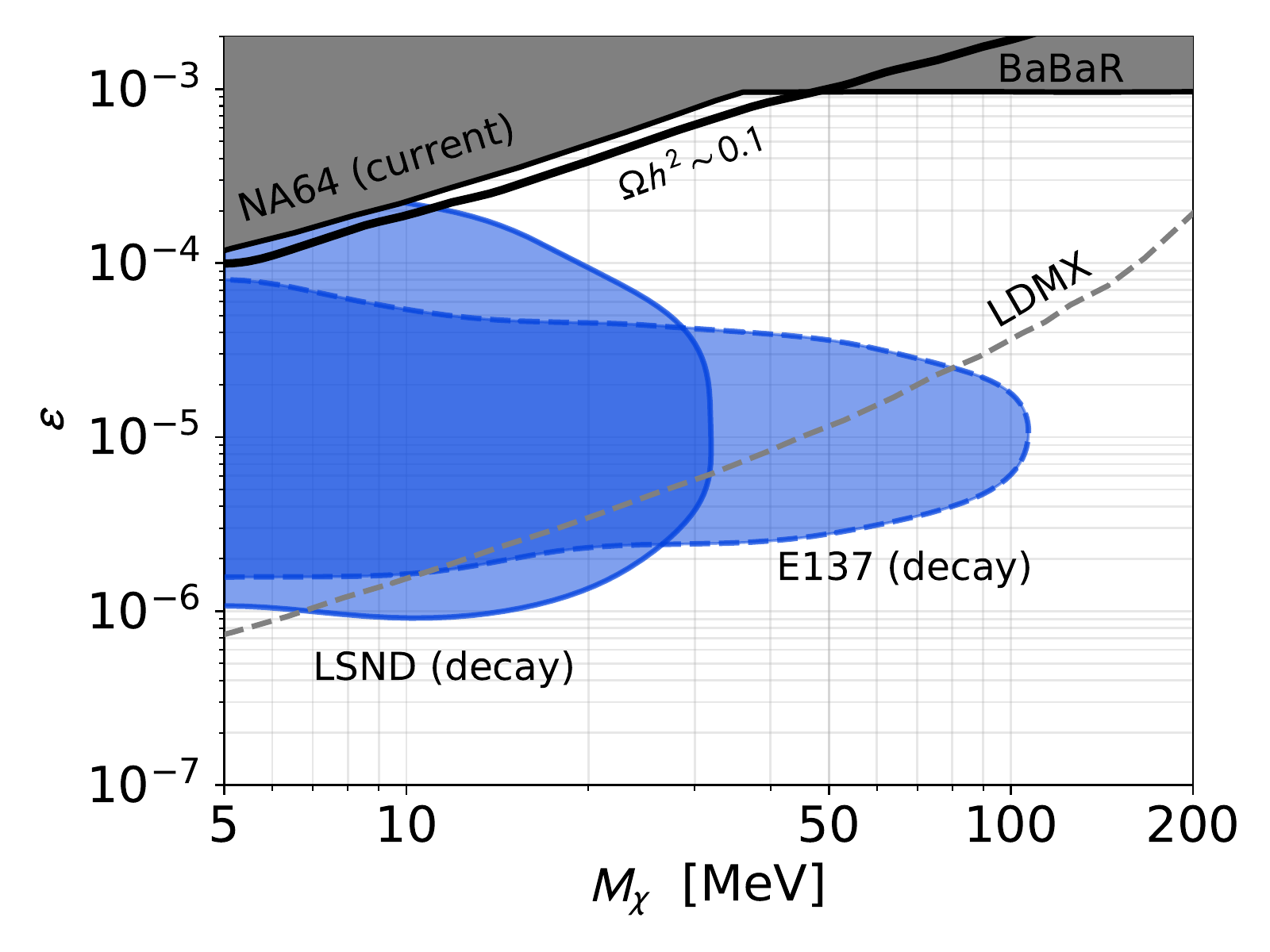}
}%
	\caption{Bounds on the Majorana DM regime from the E137~\cite{Riordan:1987aw},
	LSND~\cite{Athanassopoulos:1996ds} arising from DM scattering and $\chi_2$ decay
	(blue shaded regions) and missing energy searches (grey regions and grey dashed
	line for the LDMX projection from~\cite{Battaglieri:2017aum}). The thick black
lines correspond to points featuring the correct relic density of DM. We fixed $\azp =
0.1$, determined $\ydm$ as described in the text and varied $\mdm$ and $\eps$ using the
mass ratios   $\ms : \mdm : \mzp = 4 : 1 : 3$ and (a) $\dchi =0.8$ or (b) $\dchi =2$.}
	\label{fig:BoundmDM}
\end{figure}

\paragraph{Majorana regime (mDM).} When the mass splitting between $\chi_1$ and $\chi_2$
increases, the heavy state can decay before reaching the detector. This typically happens
for values of the kinetic mixing parameter below the missing energy search limit when
$\dchi  \gtrsim  \mdm $ for LSND and even lower values in E137 as can be seen in
Figure~\ref{fig:BoundmDM}. Consequently, the limits from, e.g., LSND present now an upper
bound for these types of process. We illustrate this in Figure~\ref{fig:BoundmDM} for two
values of the splitting parameter $\dchi = 0.8$ and $\dchi = 2$.  Interestingly,
potentially upcoming experiments such as JSNS$^2$~\cite{Ajimura:2017fld} or upgraded
SeaQuest~\cite{Berlin:2018pwi} will have a shorter beam-line and hence can reduce this
blind spot. Additionally, the typical relic density prediction is now modified with
respect to the standard iDM case and depends strongly on the effective DM coupling with
the dark photon and hence on dark Higgs sector properties as shown in
Eq.\eqref{eq:iDMcoupling} and Eq.\eqref{eq:mDMcoupling}.\footnote{For the
	Figure~\ref{fig:BoundmDM}, we have fixed $\ysl$ and $\ysr$ such that the effective Yukawa coupling $\ydm$ is given by
	\begin{align}
		y_{DM} = 	\frac{\mdm}{\mzp} \gzp \sqrt{2} \left( 1 + \frac{\dchi}{4} \right)  \ ,
\end{align} consistent with the bounds described in Appendix~\ref{sec:parameters}.}  

\begin{figure}[h!]
		%\label{fig:a}%
		\includegraphics[width=0.55\textwidth]{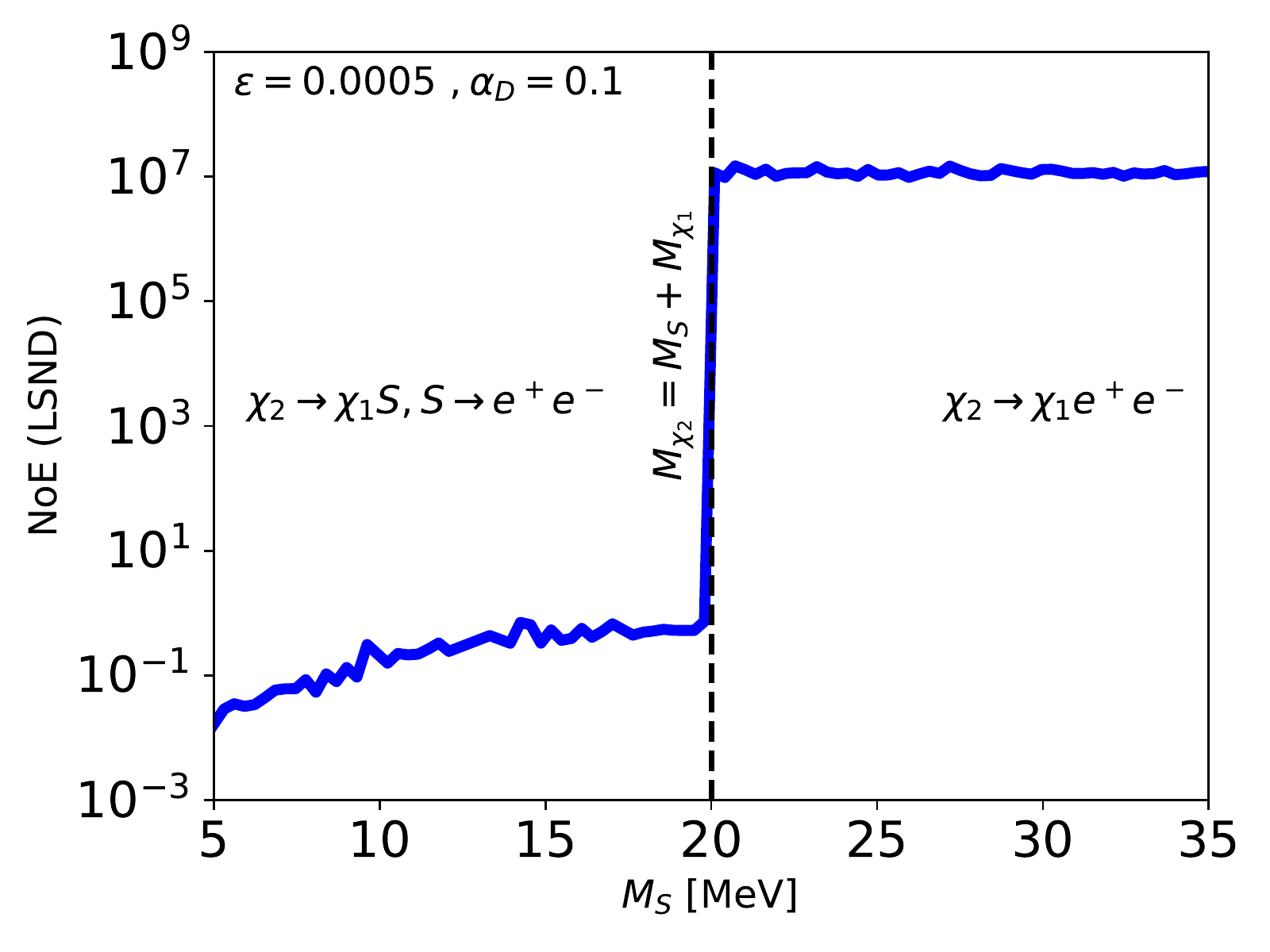}
	\caption{Expected number of events in the LSND experiment for $\eps=5 \times 10^{-4}, \azp =0.1, \mdm = 40$ MeV and $ \mdm : \mzp : \dchi= 1 : 4 : 0.5$  as function of the dark Higgs boson mass $\ms$ ($\ydm$ fixed as described in the text).}
	\label{fig:DHchain}
\end{figure}

A second interesting aspect of this regime occurs when $\ms > \dchi$. In this case, the
heavy dark sector state $\chi_2$ can decay instantaneously by emitting a dark Higgs boson.
Given that the dark Higgs boson lifetime is several orders of magnitude larger, the
expected reach is drastically reduced. We illustrate this effect in
Figure~\ref{fig:DHchain}.

\begin{figure}[t]
	\centering
	\subfloat[]{%
		%\label{fig:a}%
		\includegraphics[width=0.45\textwidth]{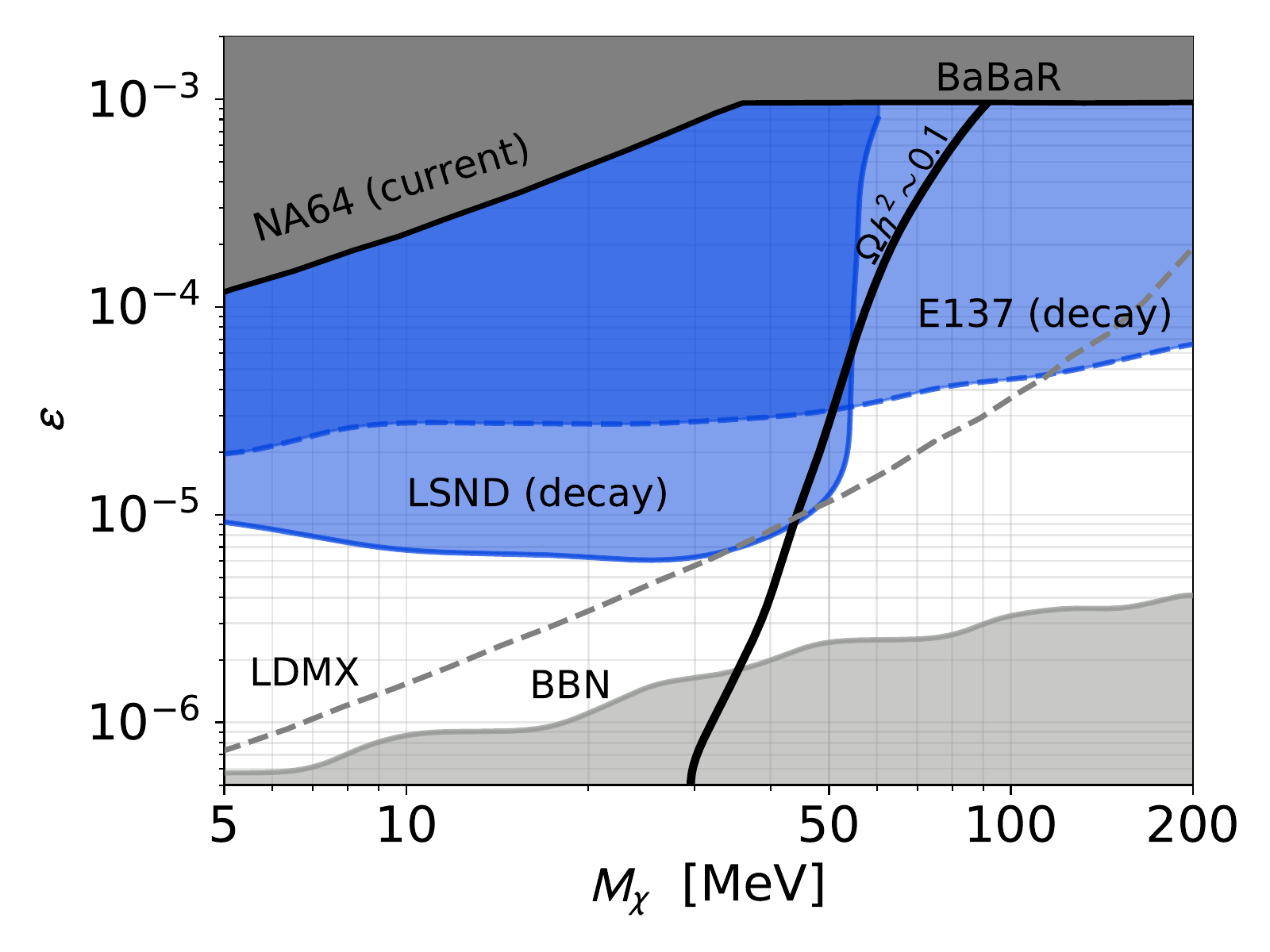}
	}%
	\hspace{0.02\textwidth}
	\subfloat[]{%
		%\label{fig:a}%
		\includegraphics[width=0.45\textwidth]{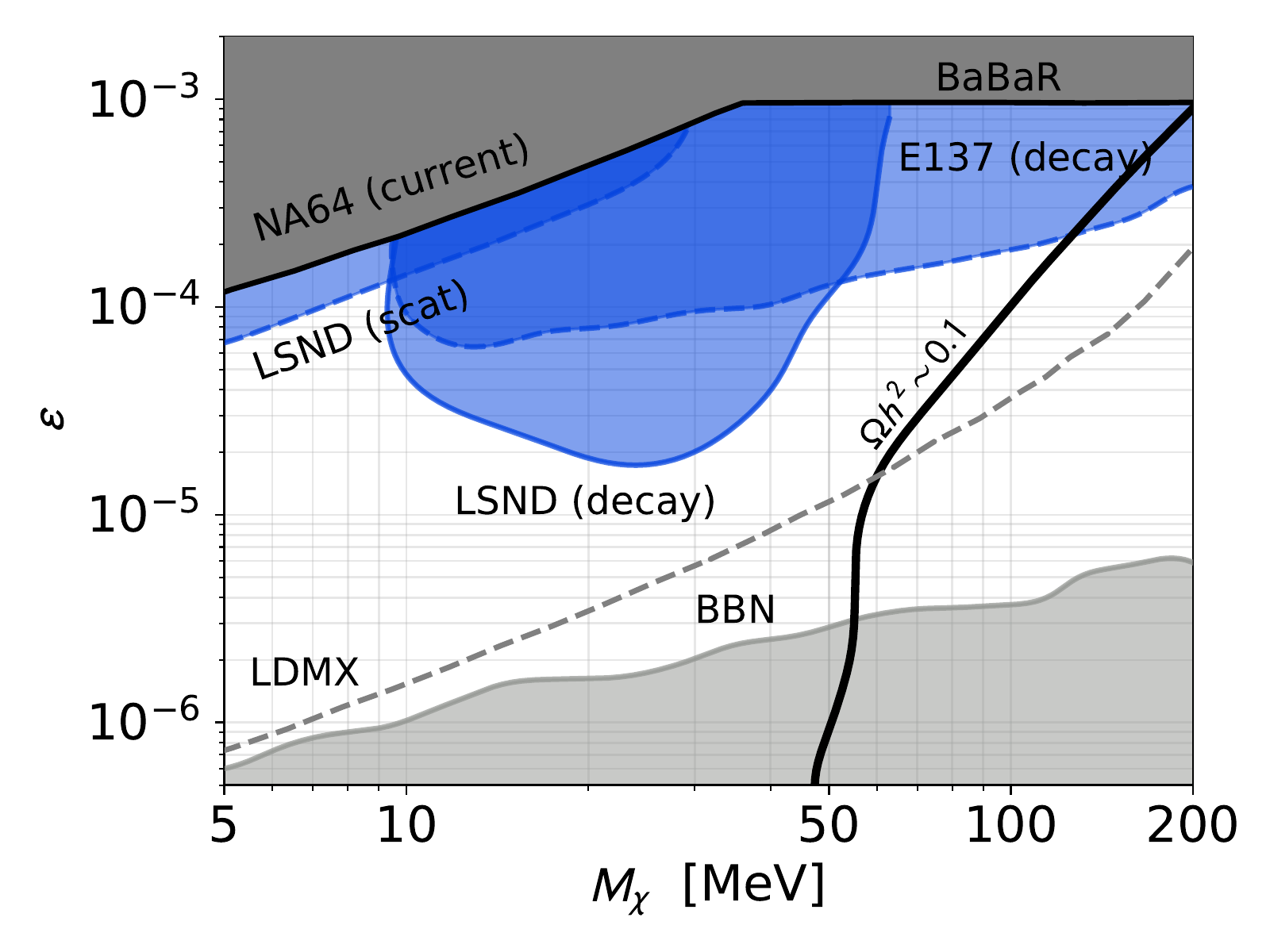}
	}%
	\caption{Bounds for the secluded regime from the E137~\cite{Riordan:1987aw}, LSND~\cite{Athanassopoulos:1996ds} arising from DM scattering and $\chi_2$ decay (blue regions) and missing energy searches (dark grey regions and grey dashed line for the LDMX projection from~\cite{Battaglieri:2017aum}). The light grey region at the bottom of the plot are excluded by BBN constraints~\cite{Darme:2017glc}. The thick black lines correspond to points featuring the correct relic density of DM. We varied $\mdm$ and $\eps$ using $\gzp =0.1$ and (a) $\ms : \mdm : \mzp :\dchi= \frac{5}{6} : 1 : \frac{5}{2} : 0.3 $  or (b) $\ms : \mdm : \mzp :\dchi= \frac{9}{10} : 1 :3 : 0.1$. The effective Yukawa coupling between DM of the dark Higgs boson is fixed at $0.005$.}
	\label{fig:secbounds}
\end{figure}

\paragraph{Secluded regime.} The accelerator bounds in this regime are very
similar to the iDM case. The main difference is the fact that for a given dark Higgs boson
mass, a lower bound on the kinetic mixing parameter is given by the BBN bounds
discussed in the previous section. We show this in Figure~\ref{fig:secbounds} for two
typical mass ratios. Note that this bound is weakened in the region of the secluded regime with $
\mdm \lesssim \ms \lesssim \mzp$ as the metastable density of dark Higgs boson is depleted
by its annihilation into DM.
Overall  this scenario is less 
constrained than the iDM or mDM ones if we assume that $\chi_1$ is the only component of DM,
mainly because lower range of $\eps$ are accessible.

\begin{figure}[t]
	\centering
	\subfloat[]{%
		%\label{fig:a}%
		\includegraphics[width=0.45\textwidth]{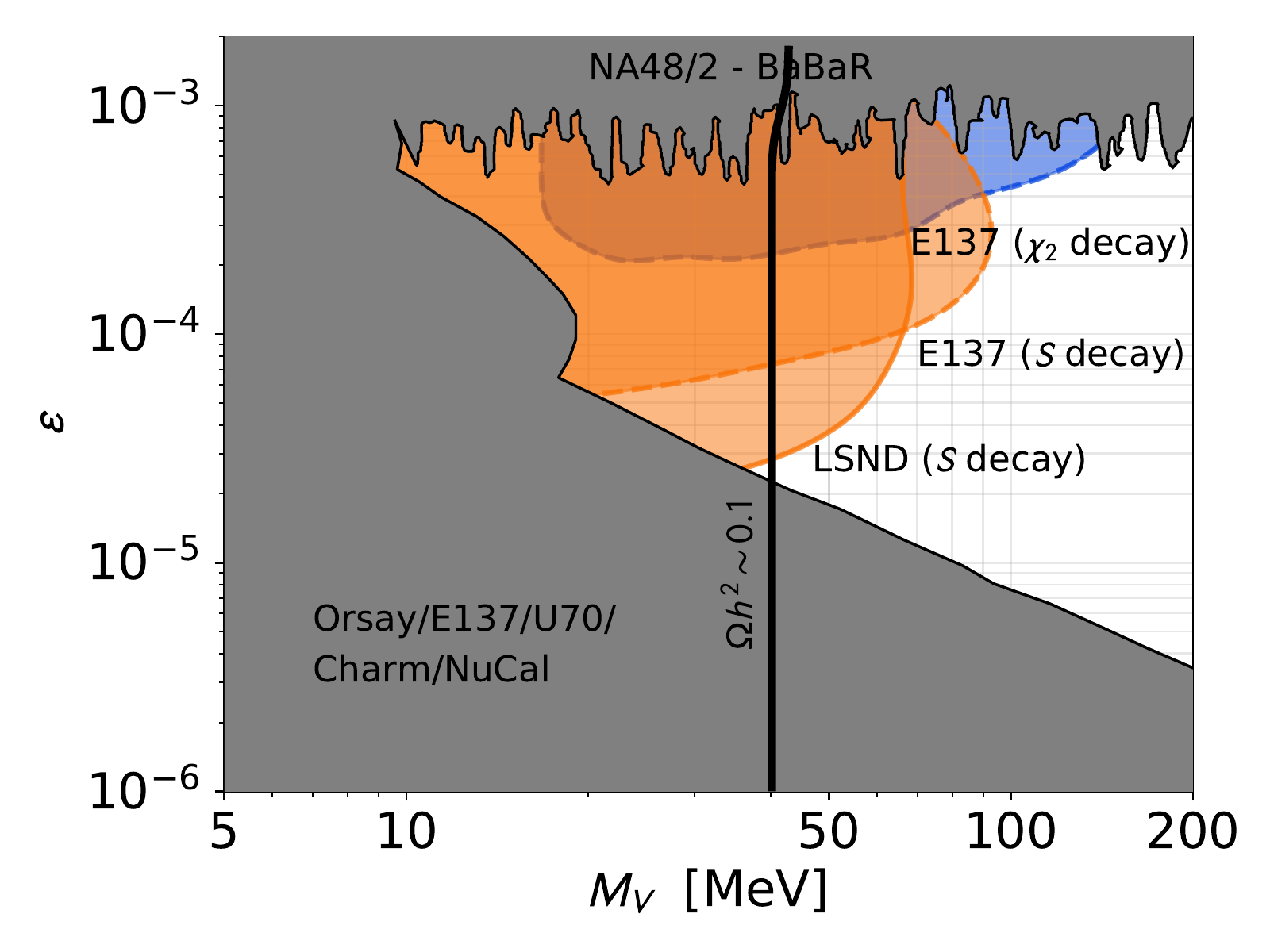}
	}%
	\hspace{0.02\textwidth}
	\subfloat[]{%
		%\label{fig:a}%
		\includegraphics[width=0.45\textwidth]{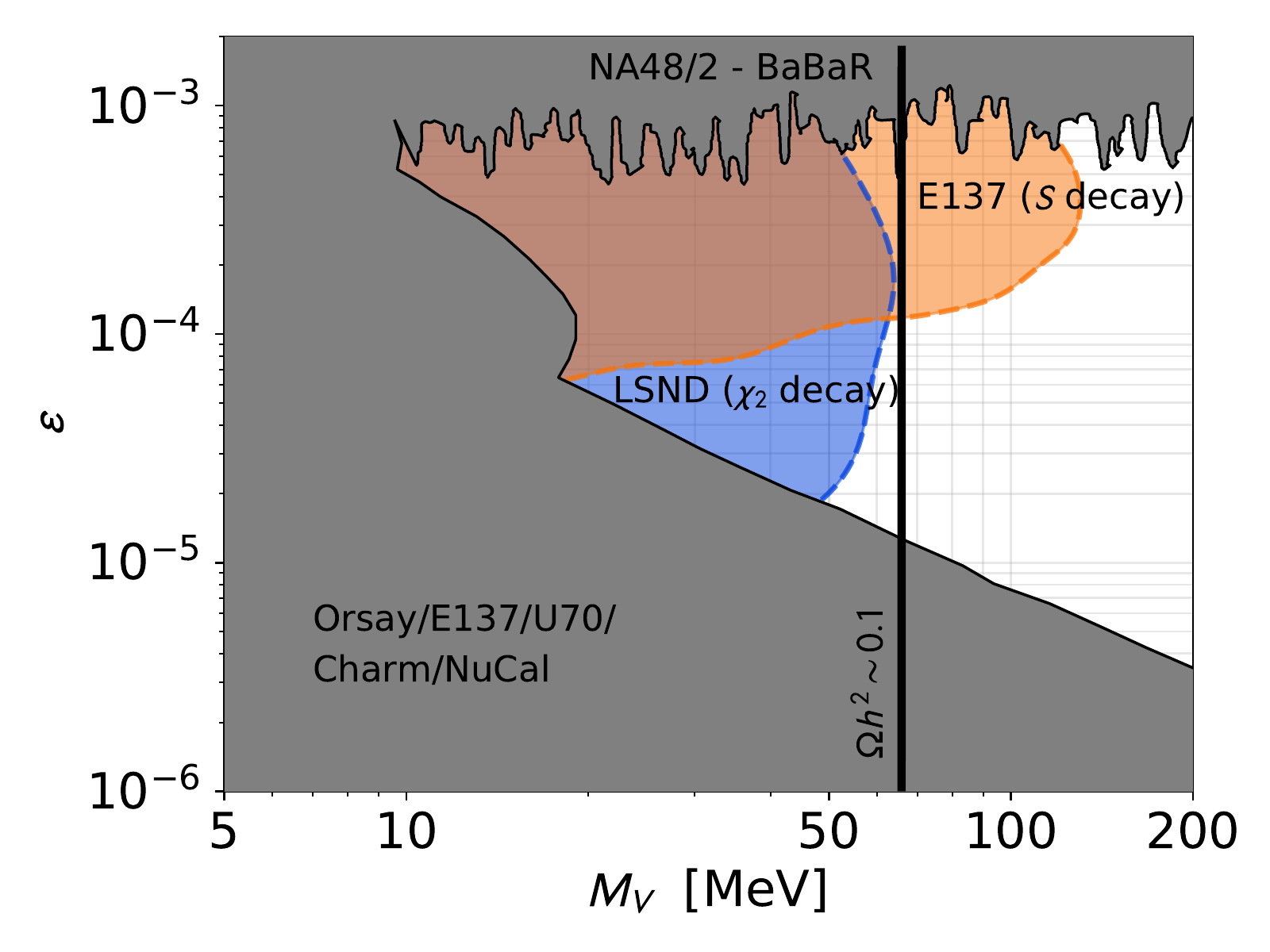}
	}%
	\caption{Bounds on the forbidden regime from the E137~\cite{Riordan:1987aw},
	LSND~\cite{Athanassopoulos:1996ds} arising from dark Higgs boson decay (orange
	shaded regions), from visible dark photon searches (dark grey regions
	from~\cite{Battaglieri:2017aum}), and from $\chi_2$ decay (blue shaded region). 
	The thick black lines correspond to points featuring the correct relic density of
	DM. We varied $\mzp$ and $\eps$ using $\gzp =0.1$ and (a) $\ms : \mdm : \mzp
	:\dchi= \frac{5}{3} : 1 : \frac{4}{3} : 0.3$ and (b) $\ms : \mdm : \mzp :\dchi=
	\frac{6}{4} : 1 : \frac{5}{4} : 0.75$.}
	\label{fig:forbbounds}
\end{figure}

\paragraph{Forbidden regime.} In this regime, the dark photon cannot decay into DM
particles. It is therefore long-lived and typically decays into electrons for the
parameter range of interest in this paper. In the absence of any other dark sector states, the
typical bounds on the dark photon decay exhibit a mass dependent gap between
$10^{-5}-10^{-3}$ (see e.g~\cite{Battaglieri:2017aum} for an up-to-date summary).
Interestingly, the presence of additional dark sector states can strongly help in closing
this blind spot. Indeed, when the dark Higgs boson is heavier than the dark photon, its
lifetime is short enough to lead to a sizable number of events in both LSND and E137
experiments. The expected reach then depends on the splitting between the dark Higgs boson
and the dark photon which control the dark Higgs boson lifetime (see
Eq.~\ref{eq:Slifetime_short}). Furthermore, both the DM and the heavy dark sector state in
general can be produced
through an off-shell dark photon, albeit at a reduced rate.  Depending on the values of
the splitting parameters $\dchi$ and $\ms-\mzp$, this can complement or increase the
reach from dark Higgs boson decays alone. Overall, dark Higgs boson-related bounds are typically stronger or equivalent to the dark sector decay ones, as we illustrate in Figure~\ref{fig:forbbounds}.

It is important to note that these bounds also applies to generic Higgsed dark photon scenario for $\ms > \mzp$ even without the
presence of DM.

\section{Summary and conclusions}\label{sec:res}

In this paper, we have considered a simple self consistent model of light (sub-GeV)
fermionic DM, whose relic density is fixed by freeze-out. This model is often
used when studying pseudo-Dirac DM since large part of its parameter space leads
to this paradigm.  Our main point is that the same model, when all particles of the
spectra are properly accounted for (in particular the dark Higgs boson), leads equally
naturally to three other DM regimes: Majorana, secluded and forbidden. The first
one corresponds to a simple Majorana DM scenario, albeit with an additional heavy
dark sector state $\chi_2$, the second relies on DM $t$-channel annihilation into
dark Higgs boson, and the last one is based on a thermally suppressed annihilation of DM into dark photons.

We have briefly reviewed the cosmological bounds on these scenarios, and then focused on
their possible signatures in accelerator-based experiments. The dark Higgs boson has been
shown to have an important role in these signatures, either indirectly through its Yukawa
couplings to the fermionic states, or directly by either leading to ``blind spots'' in
standard $\chi_2$ detection strategies or by its own decay signatures when its lifetime is
not too long. We have further considered the current bounds on the
four regimes identified stemming from the well studied results of the LSND and E137
experiments. In conclusion, we have shown that similarly to the well known pseudo-Dirac
case, the other three scenarios presented here also have bright, albeit distinct prospects
in upcoming accelerator-based experiments. 
%Future experimental searches should include these scenarios in
%their physics case. 

%\section{Sommerfeld enhancement}

%\bigskip
%%%%%%%%%%%%%%%%%%%%%%%%%%%%%%%%%%%%%%%%%%%%%%%%%%%%%%%%%%%%%%%%%%%%%%%%%%%%%%%%
%\noindent \textbf{Acknowledgments}
%\medskip
\acknowledgments

We would like to thank S. Trojanowski for helpful and interesting discussions. LD, LR and SR are supported in part by the National Science Council (NCN) research grant No.~2015-18-A-ST2-00748. LR is also supported by the project ``AstroCeNT: Particle Astrophysics Science and Technology Centre" carried out within the International Research Agendas programme of the Foundation for Polish Science co-financed by the European Union under the European Regional Development Fund. The use of the CIS computer cluster at the National Centre for Nuclear Research is gratefully acknowledged.

\newpage
\appendix

\section{Scan regions  and input parameters}
\label{sec:parameters}
Our model described in Section~\ref{sec:modbuilding} has seven free parameters: the
kinetic mixing parameter $\eps$, the dark coupling $\gzp$, the dark Higgs boson mass
$\mu_S$ and quartic coupling $\lambda_S$, the Dirac mass $m_\chi$ and the two Higgs boson Yukawa
couplings $\ysl$ and $\ysr$. Contrary to~\cite{Darme:2017glc}, we have traded the five
last parameters for more physically relevant ones, namely: the dark Higgs boson mass $\ms$
and dark photon mass $\mzp$, the DM mass $\mdm$, the splitting $\dchi \equiv
|\mhds| - |\mdone| $ between $\chi_2$ and $\chi_1$, and finally the effective Yukawa between
the DM and the dark Higgs boson $\ydm$. We report in Table~\ref{tab:input}, the range used in
the scans presented in Figure~\ref{fig:DMregions}. We nonetheless stress that we did not
consider Bayesian inference, so that Figure~\ref{fig:DMregions} should be understood only as an illustration of
the four regimes considered.

%%%%%%%%%%%%%%%%%%%%%%%%%%%%%%%%%%%%%%%%%%

\begin{table}[b]
	\begin{center}
		\makebox[\textwidth][c]{\begin{tabular}{c|c|c}
				\hline
				\hline
				\rule{0pt}{2.5ex}
			 Parameter & Range & Prior\\
				\hline
				\hline
				\rule{0pt}{2.5ex}
				$\ms$ & $5$ MeV - $1$ GeV & Log \\
				$\gzp$ & $0.01$ - $2.5$  & Log \\
				$\mzp$ & $10$ MeV- $500$ MeV  & Log \\
				$\eps$ & $0.5 \times 10^{-6}$ - $0.001$  & Log \\
				$\mdm$ & $-250$ MeV - $150$ MeV  & Linear \\
				$\dchi$ & $0.01|\mdm|$  -  $10|\mdm|$  & Log \\
				$y_{DM}$ & $-2$  -  $2$  & Linear \\
		\end{tabular}}
		\caption{Input parameters for the scans presented in Figure~\ref{fig:DMregions}.}
		\label{tab:input} 
	\end{center}
\end{table}
%%%%%%%%%%%%%%%%%%%%%%%%%%%%%%%%%%%%%%%%%%%%%%%%%%%%%%%%%%%%%%%%%

It is important to note that while this choice of variables makes it
particularly easy to relate the physics of DM annihilation to the input variables, the
price to pay is that $y_{DM}$ can only belong to a certain range. More precisely, solving
directly for $y_{DM}$ as function of the original Lagrangian parameters, we find the
accessible range to be:
\begin{align}
\label{ydmbound}
	\frac{\mdm}{\mzp} \gzp \sqrt{2} < y_{DM} <  	\frac{\mdm}{\mzp} \gzp \sqrt{2} \left( 1 + \frac{\dchi}{2} \right) \ ,
\end{align}
where the sign of the masses are important (in particular, whether or not $\mdm$ is
negative). This parameter is of course of importance to both the secluded regime, for
which we typically consider $\mdm < 0$ and $y_{DM} \ll 1$, which satisfies
Eq.~\ref{ydmbound}, and the Majorana case, for which typically $\mdm > 0$ and the
bounds on $y_{DM} $ are of importance.

\section{Monte-Carlo setup and numerical simulations}

We present in this appendix the numerical setup used to obtain the number of events in the experiments E137 and LSND, whose main characteristics are described in Table~\ref{tab:exppars}.

\begin{table}[t]
	\centering
	\begin{center}
		\makebox[\textwidth][c]{
			\begin{tabular}{l| c c c c c c}
				Name & Energy  & $\pi^0 / e^-$ produced & Target Material & Distance & Length & Area \\
				\hline
				\rule{0pt}{3ex}LSND \ \ & $0.798$ GeV  &  $0.92 \times 10^{23}$&Water/high-Z metal & $34$ m & $8.3$ m & $25.5$ m$^2$ \\
				E137 \ \ & $20$ GeV  &  $ 10^{20}$ & Al & $383$ m & $\sim 1$ m & $8$ m$^2$ \\
			\end{tabular}
		}
		\caption{Characteristics of the experiments considered. We define the detector distances from the beam target to the center of the detector. LSND has a cylindrical geometry while E137 has a square intersection with the beam axis. The effective detector distance considered for E137 is developed in the text.}
		\label{tab:exppars}
	\end{center}
\end{table}

\subsection{Production at electron and proton beam dump}

In proton beam dump experiment (and more particularly LSND), the dominant production
mechanism for a beam energy in the tens of GeV is the decay of light mesons. In this
paper, we build upon the numerical tools developed in~\cite{Darme:2017glc} for this setup and
based on a modified version of the code BdNMC~\cite{deNiverville:2016rqh}. Focusing on
LSND, we start by simulating the kinematic distribution of the light meson $\pi^0$ based on
a weighted
Burman-Smith distribution in order to account for the various target material (water, then
high-Z metal) used over the experiment lifetime. The (differential) branching ratios for
$\pi^0$ decay into the relevant final states $\chi_i \chi_j \gamma$ and $S V \gamma$ are
then calculated analytically and used to sample the kinematics of the final state and
the total number of produced events. More details can be found in~\cite{Darme:2017glc},
including the expression for the differential decay rate for $\pi^0 \rightarrow S V
\gamma$, while the rate for $\pi^0 \rightarrow \chi_i \chi_j \gamma$ is shown in
Eq~\ref{eq:diffrate}. When the heavy dark sector state's ($\chi_2$) fast decay into $S
\chi_1$ is kinematically available, we decay it by assuming the process to
occur instantaneously. Depending on the final search channel we are interested in (dark
matter scattering, $\chi_2$ decay or $S$ decay) the relevant particle kinematics are then
stored and passed to the detector simulation part of the code.

We proceed differently for electron beam dump in that we instead rely on the public code
Madgraph \cite{Alwall:2014hca} to produce the events.  In our calculation, we
use the following form factor for the target atom with an elastic and inelastic term given
by \cite{Kim:1973he,Tsai:1973py,Tsai:1986tx,Bjorken:2009mm}
\begin{align}\nonumber
	G_2^{el}&=\left(\frac{a^2 t}{1 + a^2 t}\right)^2 \left(\frac{1}{1 +
	t/d}\right)^2 Z^2\\
	G_2^{in}&=\left(\frac{{a^\prime}^2 t}{1+{a^\prime}^2 t}\right)^2 \left(\frac{1 +
		\frac{t}{4m_p^2}(\mu_p^2 - 1)}{\left(1 +
		\frac{t}{0.71\,\mathrm{GeV}^2}\right)^4}\right)^2 Z
	\label{ff}
\end{align}
where $a=111Z^{-1/3}/m_e$ with $m_e$ being the electron mass, $d=0.164$ GeV$^2A^{-2/3}$
with $A$ being the number of nucleons in the target, $a^\prime=773Z^{-2/3}/m_e$ with $Z$
being the number of protons in the target, $m_p$ is the proton mass and $\mu_p=2.79$.  The
atom-dark photon vertex then becomes $(P_i+P_f)_\mu\,\sqrt{G_2}$, where $P_i$ and $P_f$ are
initial and final momenta of the atom respectively, while $G_2=G_2^{el}+G_2^{in}$.  We
implement the above atomic and nuclear form factors using the in-built procedure of
Madgraph \cite{madff}.
%\LD{Add description of the form factors and Madgraph part}.

In order to account for the energy damping of the electron beam in E137 as it goes through
the aluminum target, we run the previous code for three different energies of $18$ GeV,
$10.5$ GeV and $3.5$ GeV. The resulting cross-section is then weighted by the usual
energy distribution from Ref.~\cite{Tsai:1986tx}:
\begin{align}
	I(E_0,E_e) = \frac{E_0}{E_e} \int_0^\infty ds \frac{1}{E_o} \frac{\left[\ln (E_0/E_e) \right]^{4 s/3  -1}}{\Gamma [4 s/3]} \ ,
\end{align}
where we have directly integrated over the thick target length (approximated to infinity),
$E_0 = 20$ GeV and $E_e = 18, 10.5, 3.5$ GeV in our case.\footnote{Importantly,
when accounting for the energy damping, we did not include the transverse broadening
of the beam as it looses energy in the target. This should further reduce the number of
events in the low energy bin. However, the high energy threshold of the E137 experiment
implies that most of the events from this bin do not eventually pass the detector cuts, so
that including this effect should not modify our result significantly.} Finally we
reconstruct an unweighted sample of final states by adding events from our three energy
bins proportionally to the bin's weight. This unweighted sample of final states' kinematic
is then passed to the detector simulation part of the code, along with the total number of
produced events. 

\subsection{Decay and detector simulations}

This part of our simulation, also loosely based on BdNMC~\cite{deNiverville:2016rqh},
takes as input the kinematics information of final states and the total number of
produced events (regardless of the electron or proton beam dump origin of the events). The
scattering detection simulation is imported from the original BdNMC code and is described
in~\cite{deNiverville:2016rqh}. In this section, we will focus on the detector simulation
for a decaying $\chi_2$ or $S$ particles.

We start by analytically calculating the lifetime of the particle and its decay probability within a given decay volume consisting of the detector itself for LSND and a fraction of the open space before the detector for E137 (we will comment on this particular aspect at the end of this section). We then sample the decay product kinematics using the hard-coded expression for the differential decay rate for the processes $\chi_2 \rightarrow \chi_1 e^+ e^-$, $S \rightarrow e^+ e^-$ or $S \rightarrow V e^+ e^-$. At this point we determine whether any more decay occurs before the detector (e.g dark photon decay for the $S \rightarrow V e^+ e^-$ case) then pass the decay products to the detector simulation.

First, the simulation of the LSND detector response to the event is based on the
search in \cite{Aguilar:2001ty}, so that we require the electrons in final state to either be
reconstructed as a single track (namely with an angular spread of the $e^+ e^-$ pair below
$12^{\circ}$) or that only one of the electron is reconstructed (using an electron
detection efficiency of $19\%$). Overall, the reconstructed object is then required to
have an energy between $18$ and $50$ MeV and be forward-oriented with $\cos \theta_b >
0.9$, where $\theta_b$ is the angle with the beam-line in the laboratory frame, in order
to fake the signature of an elastic events as searched for in~\cite{Aguilar:2001ty}.
Following~\cite{Izaguirre:2017bqb}, we then use a 55-events limit to derive the bounds from
this experiment.

Second, for the E137 experiment, we require that at least one of the electrons crosses the
detector with an angle to the beam line smaller than $30$ mrad and an energy $E > 1$ GeV.
No such events were observed by E137 (see, e.g Figure 9 of Ref.~\cite{Bjorken:1988as}) so
that we place an upper limit of 3-events. An important comment regarding E137 is that the decay
volume in front of the detector was an open space at atmospheric pressure. The typical
radiation length for electrons in the air is of $304$m~\cite{Patrignani:2016xqp}, but
perhaps more importantly the corresponding Moliere radius is $73$m. This implies that after a
distance of order meter, the energy of the electromagnetic shower is spread out in a transverse
direction significantly larger than the spread of an event as seen by the E137 (see Figure
8 of Ref.~\cite{Bjorken:1988as}). While a proper modeling of the electromagnetic shower
will be probably needed to go beyond the order of magnitude evaluation, we will take the
simpler approach of considering only those events for which the decay occurs within $\sim
1$ m of the detector.

\bibliographystyle{utphys}
\bibliography{FDM}

\end{document}